\documentclass[a4paper]{IEEEtran}
\usepackage{amsmath}
\usepackage{algorithm,algorithmic}
\usepackage[pdftex]{graphicx}
\usepackage{amsmath,amssymb}
\usepackage{bm}
\usepackage{color}

\newtheorem{corollary}{Corollary}
\newtheorem{theorem}{Theorem}

\title{Ordinary Differential Equation-based \\Sparse Signal Recovery}

\author{%
  \IEEEauthorblockN{
  		Tadashi Wadayama and Ayano Nakai-Kasai}\\
  \IEEEauthorblockA{\IEEEauthorrefmark{1}%
		Nagoya Institute of Technology,
		Gokiso, Nagoya, Aichi 466-8555, Japan,\\
 		wadayama@nitech.ac.jp, nakai.ayano@nitech.ac.jp}
 		\thanks{
Part of this research was presented 
at the IEICE International Symposium on Information Theory and Its Applications 2022 (ISITA2022)~\cite{Wadayama2022}.
}
}
\begin{document}
\maketitle

\begin{abstract}
This study investigates the use of continuous-time dynamical systems 
for sparse signal recovery. The proposed dynamical system is in the form of a 
nonlinear ordinary differential equation (ODE) derived from 
the gradient flow of the Lasso objective function. 
The sparse signal recovery process of this ODE-based approach 
is demonstrated by numerical simulations using the Euler method.
The state of the continuous-time dynamical system eventually converges 
to the equilibrium point corresponding to the minimum of the objective 
function.
To gain insight into the local convergence properties of the system, 
a linear approximation around the equilibrium point is applied, 
yielding a closed-form error evolution ODE. 
This analysis shows the behavior of convergence to the equilibrium point. 
In addition, a variational optimization problem is proposed to optimize 
a time-dependent regularization parameter in order 
to improve both convergence speed
and solution quality. 
The deep unfolded-variational optimization method is introduced as a means 
of solving this optimization problem, 
and its effectiveness is validated through numerical experiments.
\end{abstract}

\section{Introduction}

Continuous-time dynamical systems are ubiquitously used in signal processing.
Electronic circuits are a common example where the behavior is represented 
by nonlinear ordinary differential equations (ODEs).
These circuits typically handle low-dimensional signals and simple signal processing tasks. 
From both a practical and theoretical point of view,
continuous-time systems offer a number of advantages when processing high-dimensional signals. 
One such advantage is the potential for analog-domain implementations of 
high-dimensional signal processing tasks
that are typically performed in the digital domain. 

Recently, the trend in Moore's law, which asserts that the number of transistors 
in an integrated circuit will double every 24 months, 
has not been reflected in reality. 
A major limiting factor is the power consumption of the integrated circuit, 
which hampers the ability to handle large-scale signal processing problems 
that require high throughput. 
For instance, base stations (BSs) 
in wireless network systems~\cite{MUMIMO}\cite{Yang} need to perform 
a significant number of signal processing tasks, 
including multiple-input and multiple-output (MIMO) 
signal detection and beamforming. In next-generation systems (e.g., those 
beyond 5G or 6G), the limitations of the central processing unit (CPU) 
are likely to become a major obstacle 
to achieving desired specifications. 
A potential solution to this challenge would be to use an analog-domain 
signal processing unit as a specialized processing unit with lower power consumption.

Analog computing~\cite{Ulmann}\cite{Haensch}\cite{Welser} can be effectively employed in a variety of domains, including the electrical and optical domains. 
An analog computer capable of simulating any nonlinear ODE 
is composed of analog adders, multipliers, integrators, and other nonlinear devices. 
Operational amplifiers can be used to implement such analog devices required for analog computing. 
Prior studies have been conducted on the use of specialized analog domain processing units~\cite{Huang, Guo}
based on these electrical devices to achieve better energy efficiency. 
In the field of optical computing, optical adders and multipliers 
have been used to solve ODEs~\cite{LU}. 
Recent examples of optical computing include a neural network implemented in the optical domain~\cite{Zhang2021}
and an optical Ising machine~\cite{Prabhu} for combinatorial optimization.
Programmable integrated optical circuits~\cite{Capmany} 
are becoming an active research topic in optical computing.
While it is too early to make a definitive statement, it appears that analog computing 
has the potential to perform large-scale signal processing tasks
with high energy efficiency and extremely fast computation in the future.

Another advantage of continuous-time dynamical systems is that 
they can provide deeper insights into discrete-time algorithms, 
which are the counterpart of continuous-time systems. 
They also offer a way to generate new discrete-time algorithms. 
A numerical method for solving ODEs, such as the Euler method or the Runge-Kutta method~\cite{NumericalMethods}, 
can be used to discretize a high-dimensional ODE corresponding to a signal processing task. This principle has been proposed in the study of neural ODEs~\cite{Chen}. The discrete-time algorithms derived in such a manner inherit properties of the original ODE. This correspondence between an ODE and a discrete-time algorithm opens an insightful perspective for analyzing discrete-time algorithms.

There have been several studies into 
{\em ODE-based signal processing}, 
including a recent work~\cite{Nakai} 
that proposed an ODE-based method for MMSE signal detection in MIMO systems 
and presented a concise analytical formula for the mean squared error. 
This work~\cite{Nakai} deals with a linear ODE that is relatively easy to analyze. 
Despite the usefulness, 
there remains a plethora of opportunities for further research 
in {\em nonlinear} ODE-based signal processing.

The focus of this paper is on an ODE-based method for 
{\em sparse signal recovery}~\cite{CS1,CS2}. 
Our work is motivated by the fact that most tasks in sparse signal recovery (or compressed sensing)   
are high-dimensional 
signal processing tasks that are not trivial to solve. 
Furthermore, there is a growing demand for high-speed and large-scale 
sparse signal recovery tasks in a compressed sensing context~\cite{Alg_survey}. 

A number of discrete time sparse signal recovery algorithms~\cite{Alg_survey}
have been developed based on the Lasso formulation~\cite{LASSO, MP,LARS, CDA}.
The Iterative Shrinkage Thresholding Algorithm (ISTA)~\cite{ISTA, ISTA2}
is one of the best-known algorithms for solving the Lasso problem. 
ISTA is an iterative algorithm comprising two processes: 
a linear estimation process and a shrinkage process based on 
a soft shrinkage function.
ISTA can be seen as a proximal gradient descent algorithm~\cite{Prox}
and can be directly derived from the Lasso formulation.
Another powerful iterative algorithm is Approximate Message Passing (AMP)~\cite{reBP,AMP} which provides much faster convergence than ISTA. 
Ma and Ping proposed Orthogonal AMP (OAMP)~\cite{OAMP},
which can handle various classes of sensing matrices 
including unitary invariant matrices. 
Rangan et al. proposed Vector AMP~\cite{VAMP} 
for right-rotationally invariant matrices and provided a theoretical justification for its state evolution.

Although the advancement of these sparse signal recovery 
algorithms is significant, all of them
are discrete-time algorithms. 
The extension to continuous-time signal processing 
is not straightforward.
Our approach is based on the Lasso formulation~\cite{LASSO}, 
which can be viewed as 
a typical regularized least-squares (LS) objective function. 
The proposed ODE corresponds to 
the {\em gradient flow} based on the Lasso objective function.
To gain insight into the local convergence properties of the system, 
a linear approximation around the equilibrium point is applied,
which yields a closed-form error evolution ODE. 
The analysis tracks system behavior as the solution converges to the equilibrium point. 
In addition, a variational optimization problem is proposed to optimize 
a time-dependent regularization parameter in order to improve both convergence speed
and solution quality. 
A {\em deep unfolding}-based method is presented for solving the variational problem.
The methodology presented in the paper is directly applicable 
to any type of regularized LS problem. 

The remainder of the paper is organized as follows:
In Section \ref{sec:preliminaries}, the system model and the Lasso formulation used in the paper are introduced.
Section \ref{sec:ODE-based} presents the proposed ODE-based method for sparse signal recovery and includes numerical demonstrations.
In Section \ref{sec:localconveregnce}, an analysis of the local convergence of the proposed ODE-based method is given, with the key idea being the use of a linear approximation around the equilibrium point. Numerical examples are provided to support the theory presented in this section.
In Section \ref{sec:DU}, a parametric ODE, which includes a continuous-time function to be optimized, is introduced. To improve the convergence behavior and solution quality, a variational optimization problem including the parametric ODE must be solved. A deep unfolding-based optimization method is presented for solving this problem.
Section \ref{sec:conclusion} concludes the discussion.

\section{Preliminaries} 
\label{sec:preliminaries}

\subsection{Notation}
The following mathematical notation is used throughout the paper:
The symbols $\mathbb{R}$ and $\mathbb{R}_+$ represent the set of real numbers and the set of positive real numbers, respectively.
The one-dimensional Gaussian distribution with mean $\mu$ and variance $\sigma^2$ is denoted by ${\cal N}(\mu, \sigma^2)$.
The multivariate Gaussian distribution with mean vector $\bm \mu$ and covariance matrix $\bm \Sigma$ is represented by ${\cal N}(\bm \mu, \bm \Sigma)$.
Assume that a function $f:\mathbb{R} \rightarrow \mathbb{R}$ is given.
The same function $f$ can be applied to $\bm x \equiv (x_1,x_2,\ldots, x_n)$ as $f(\bm x) = (f(x_1), f(x_2),\ldots, f(x_n))$.
The expectation operator is denoted by ${\sf E}[\cdot]$.
The notation $\mbox{diag}(\bm x)$ denotes the diagonal matrix 
whose diagonal elements are given by $\bm x \in \mathbb{R}^n$.
The matrix exponential $\exp(\bm X) (\bm X \in \mathbb{R}^{n \times n})$ is defined by
\begin{align}
\exp(\bm X) \equiv \sum_{k=0}^\infty \frac{1}{k!}\bm X^k.	
\end{align}
The spectral norm of $\bm X\in \mathbb{R}^{n \times n}$
is denoted by $\|\bm X\|_2$.
The notation $[n]$ denotes the set of consecutive integers
from $1$ to $n$. 

\subsection{System model}

In this paper, we follow a common system model for compressed sensing~\cite{CS1,CS2}. A sparse vector $\bm s \in \mathbb{R}^n$ is given, and it is assumed that the original sparse signal follows the Bernoulli-Gaussian distribution, where a non-zero element follows a Bernoulli distribution with probability $p$ and a non-zero element follows a Gaussian distribution ${\cal N}(0,1)$. A linear observation vector $\bm y \in \mathbb{R}^m$ is generated by
\begin{align} \label{cs_model}
\bm y = \bm A \bm s + \bm n,
\end{align}
where $\bm A \in \mathbb{R}^{m \times n}$ is the sensing matrix. The vector $\bm n \in \mathbb{R}^m$ is an additive Gaussian noise vector following $\bm n \sim {\cal N}(\bm 0, \sigma^2\bm I)$. The goal of the sparse signal recovery problem is to estimate the original sparse vector $\bm s$ from the knowledge of $\bm y$ and $\bm A$ as accurately as possible.

\subsection{Lasso formulation of sparse signal recovery}

The Lasso objective function is defined as
\begin{align} \label{lasso}
	f(\bm x) \equiv \|\bm y - \bm A \bm x\|^2_2 + \lambda \|\bm x\|_1,
\end{align}
where $\lambda ( > 0)$ is the regularization parameter. The L1 regularization term is included in order to promote a sparse solution in the regularized LS estimation. The optimization problem is then defined as
\begin{align}
	\hat{\bm x} = \mbox{arg min}_{\bm x \in \mathbb{R}^n} \ f(\bm x).	
\end{align}
This optimization problem is convex and can be solved by any convex optimization algorithm. A common approach to solving the Lasso problem is to use a proximal gradient descent method, 
such as ISTA~\cite{ISTA2}.

\section{ODE-Based Sparse Signal Recovery}
\label{sec:ODE-based}

\subsection{Gradient flow for sparse signal recovery}

In the following argument, we introduce a gradient flow~\cite{Strogatz} 
based on the Lasso objective function.

Gradient flow is a dynamical system of the form
\begin{align}
	\frac{d \bm x(t)}{dt} = -\nabla {\cal E}(\bm x(t)),
\end{align}
where ${\cal E}$ is an energy function.
The value of the energy ${\cal E}(\bm x(t))$ decreases monotonically 
as time $t$ increases. The behavior of many physical systems can 
be described by an energy minimization process.

The Lasso objective function $f(\bm x)$ in (\ref{lasso}) is not 
differentiable at $\bm x$ if $\bm x$ includes zero elements. 
This means that the gradient $\nabla f(\bm x)$ is not well-defined at some $\bm x$.
Instead of the objective function $f$ defined in (\ref{lasso}), 
we here use 
the differentiable function $g: \mathbb{R}^n \rightarrow \mathbb{R}$ defined by
\begin{align} \label{def_g}
	g(\bm x) \equiv \|\bm y - \bm A \bm x\|_2^2 + \lambda \xi_\alpha(\bm x)
\end{align}
as an energy function in the gradient flow. 

\begin{figure}
\centering
\includegraphics[width=\columnwidth]{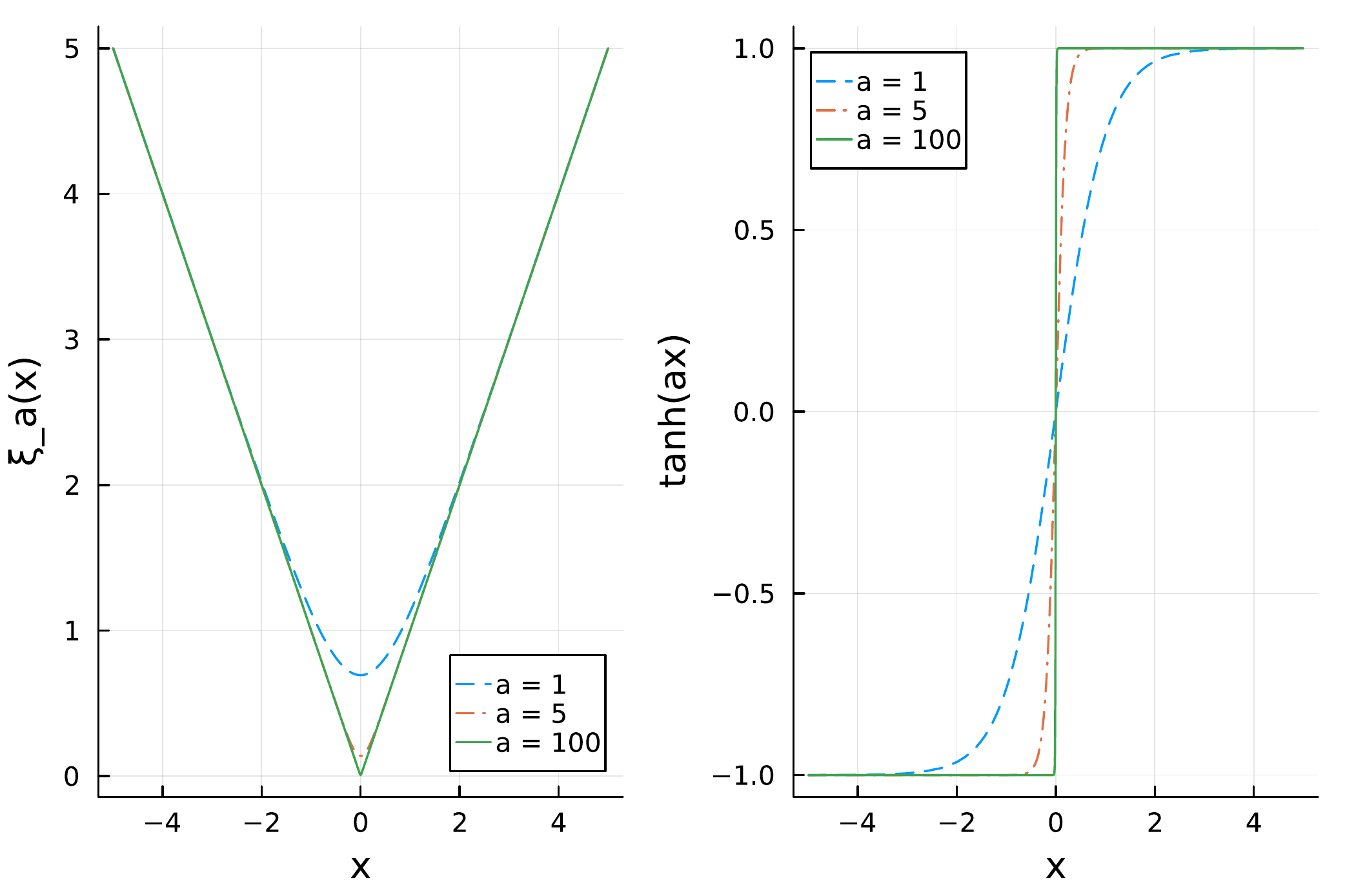}	
\caption{Proxy function for $|x|$: Left: $\xi_\alpha(x)$, Right: $\xi_\alpha'(x) = \tanh(\alpha x)$}
\label{xi_fig}
\end{figure}

The function $\xi_\alpha:\mathbb{R}\rightarrow \mathbb{R}$ is given by
\begin{align}
	\xi_\alpha(x) \equiv \frac{\log(\exp(\alpha x) + \exp(-\alpha x))}{\alpha},
\end{align}
where $\alpha \in \mathbb{R}_+$ is called the proximity parameter.
The function $\xi_\alpha(x)$ can be seen as a proxy function of the absolute value function,
i.e., $\xi_\alpha(x) \simeq |x|$ for sufficiently large $\alpha$.
At the limit $\alpha \rightarrow \infty$, the function $g$ converges to the
function $f$.
The derivative function of $\xi_\alpha$ is given by
\begin{align}
	\xi_\alpha'(x) = \tanh(\alpha x).
\end{align}
Figure \ref{xi_fig} shows the shapes of 
$\xi_\alpha(x)$ and $\xi_\alpha'(x) = \tanh(\alpha x)$
for $\alpha = 1,5,100$. It can be observed that
$\xi_\alpha(x)$ approaches the absolute value function
$|x|$ as $\alpha$ increases.

Note that we thus have the gradient of the energy function:
\begin{align} \label{grad_g}
	\nabla g(\bm x) = \bm A^T (\bm A \bm x(t) - \bm y) + \lambda \tanh(\alpha \bm x(t)).
\end{align}
In this paper, we focus on the gradient flow
defined by the 
following nonlinear ODE:
\begin{align} \label{diff_eq}
	\frac{d \bm x(t)}{dt} 
	&= - \left(\bm A^T (\bm A \bm x(t) - \bm y) + \lambda \tanh(\alpha \bm x(t)) \right),\\
	\bm x(0) &= \bm x_0,
\end{align}
where $\bm x(t) \in \mathbb{R}^n$.
The variable $t$ represents continuous time in our context.
The vector $\bm x_0$ is the initial state providing 
a boundary condition.
Because the above ODE contains a nonlinear term 
$\tanh(\alpha \bm x(t))$, it is not possible to obtain a succinct closed-form solution.
Therefore, in order to understand the dynamics of the system, 
we must resort to using numerical methods to solve the differential equation.

\subsection{Numerical solution via Euler method}
\label{Euler_sec}

The Euler method is the simplest numerical method 
for solving simultaneous nonlinear differential equations~\cite{NumericalMethods}.
Although the convergence order of the Euler method is inferior to 
that of higher-order methods such as the Runge-Kutta methods~\cite{NumericalMethods}, 
the Euler method is simple to use and can provide sufficiently precise solutions 
if sufficient discretization of the time interval is used.
Thus, in this work, we use the Euler method to solve (\ref{diff_eq}). 

Consider the ODE in (\ref{diff_eq}).
Assume that we need an approximation of the numerical solution of the above ODE
in the time interval $0 \le t \le T$.
This interval is first divided into $N$ bins.  
The discrete-time ticks
$
t_{k} = k \eta (k = 0, 1, \ldots, N)
$
define the boundaries of the bins, 
where the width of a bin, $\eta$, is given by 
\begin{align}
\eta \equiv \frac{T}{N}.		
\end{align}
It should be noted that the choice of the bin width  $\eta$ 
is crucial in order to ensure the stability and the accuracy of the Euler method. 
A small width leads to a more accurate solution, but requires more computational time. 
A large width may be computationally efficient but may lead to instability in the solution.
Let us define a discretized sample $\bm x^{(k)}$ as $\bm x^{(k)} \equiv \bm x(t_k)$.

By using the Euler method,
the solution of (\ref{diff_eq}) can be approximated by
the following recursive formula:
\begin{align} 
\bm x^{(k+1)} 
&= \bm x^{(k)}	- \eta \nabla g(\bm x^{(k)}) \\
&= \bm x^{(k)}	- \eta \left(\bm A^T (\bm A x^{(k)} - \bm y) + \lambda \tanh(\alpha \bm x^{(k)}) \right), \\
\bm x^{(0)} &= \bm x_0,
\end{align}
for $k=0,1,2,\ldots, N$.
The initial value is set to $\bm x^{(0)} = \bm x_0$.
\subsection{Numerical experiments}

To confirm the behavior of the ODE-based sparse	signal recovery (\ref{diff_eq}), 
numerical experiments were conducted.
The parameter settings for the experiments were as follows.
The length of the sparse signal $\bm s$ was set to $n = 128$. 
The length of the observation vector $\bm y$ was $m = 64$.
The standard deviation of the Gaussian noises was set to $\sigma = 0.1$.
The sensing matrix $\bm A$ was 
randomly generated. Each element in $\bm A$ follows ${\cal N}(0,1)$. In the following part of this paper, 
the same $\bm A$ is used for the experiments.

\begin{figure}[htbp]
\begin{center}
\includegraphics[width=\columnwidth]{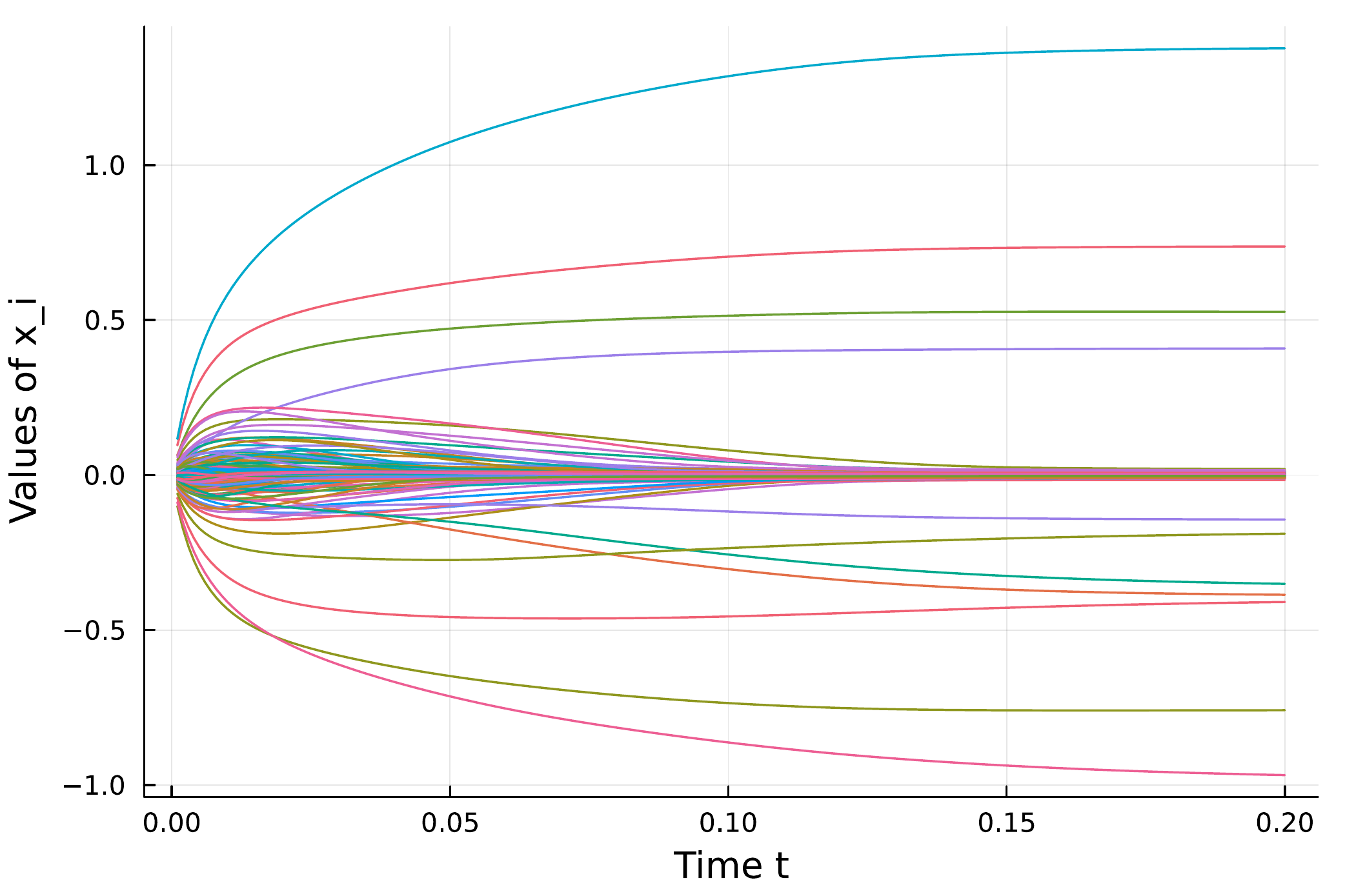}
\caption{A solution of the ODE (\ref{diff_eq}). Each component of $\bm x(t)$
is depicted as a function of time $t$. The parameters are:
$n=128, m = 64, p=0.1, \sigma = 0.1, \lambda = 5, \alpha = 50$, $\bm x_0 = \bm 0$, $T=0.5$.}
\label{trajctory}
\end{center}
\end{figure}

Figure \ref{trajctory} shows the state trajectories 
of $\bm x(t)$, which can be considered as a solution of
the ODE (\ref{diff_eq}). Each component of 
$\bm x(t)$ corresponds a curve in Fig.~\ref{trajctory}.
The Euler method with $N = 500$ was used to solve the ODE.
The observation vector $\bm y$ was generated randomly 
according to the system model (\ref{cs_model}).
We can see that the values of many components 
approach zero and that only a faction of the curves corresponding to non-zero elements in $\bm s$ deviate from zero. 
Namely, we can see that $\bm x(t)$ becomes a sparse vector with time.

\begin{figure}[htbp]
\begin{center}
\includegraphics[width=\columnwidth]{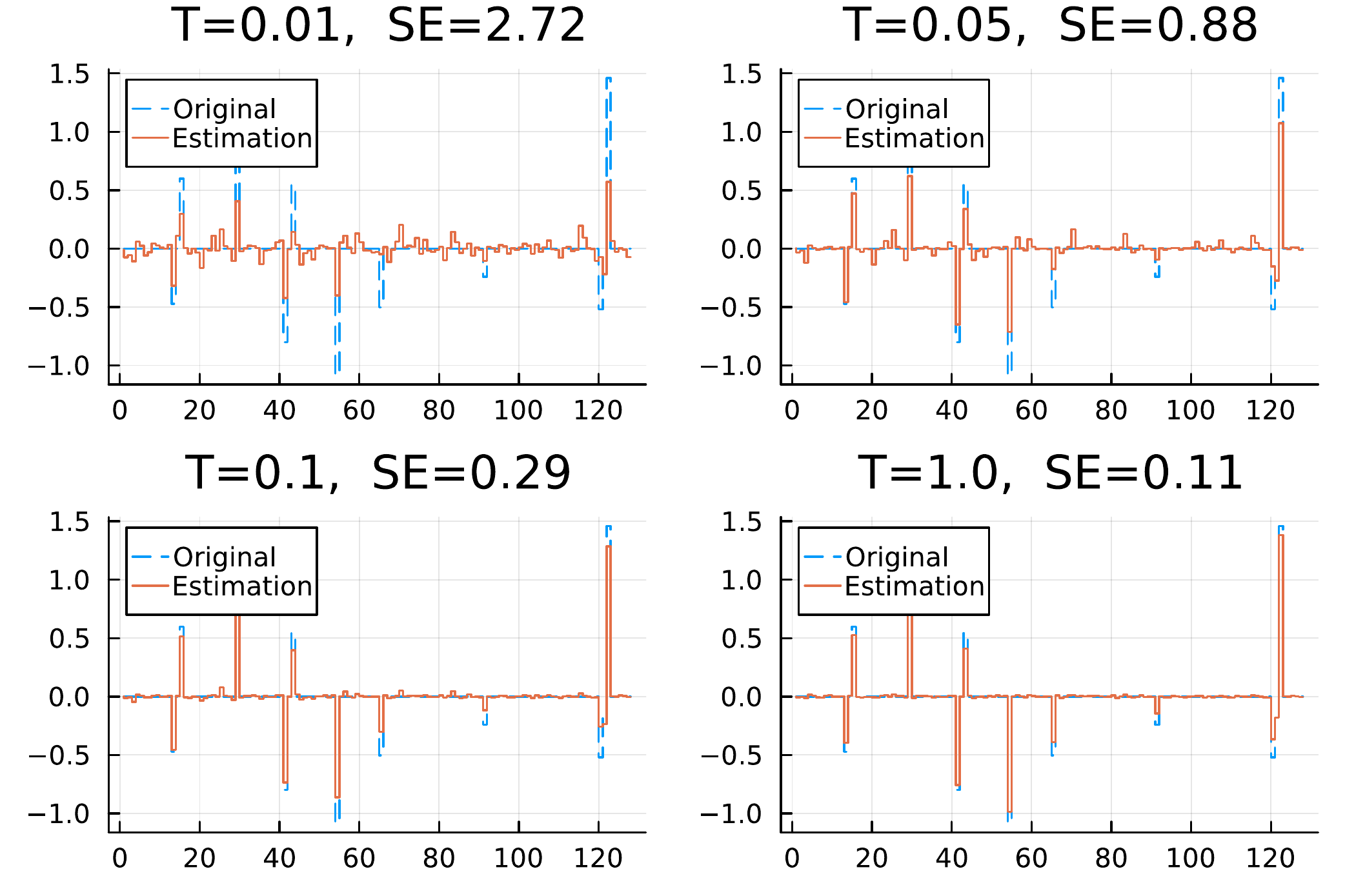}
\caption{Example of the sparse signal recovery process. The horizontal axis represents the index
of a vector of length 128. The value of $x_i(T)$ is plotted at the $i$th index position.
The parameters are:
$n=128, m = 64, p=0.1, \sigma = 0.1, \lambda = 5, \alpha = 50$, $\bm x_0 = \bm 0$.}
\label{reconst}
\end{center}
\end{figure}

Figure \ref{reconst} shows sparse signal recovery processes based on
ODE (\ref{diff_eq}) with regularization parameter $\lambda = 5$ and 
proximity parameter $\alpha = 50$.
For the numerical evaluation, the Euler method with $N = 5000$ and $T=0.5$ was used 
for each $T \in \{0.01, 0.05, 0.1, 1.0\}$.
The squared error $\|\bm s - \bm x(T)\|_2^2$ is represented as SE in 
Fig.~\ref{reconst}.
At $T = 0.01$, the reconstructed vector appears to be a noise vector.
However, we can observe that the reconstructed vector 
approaches the original signal as $T$ increases.  
These results confirm that the ODE-based sparse signal recovery can be successful with 
appropriate parameter settings.

To confirm the accuracy of the Euler method,
we conducted an experiment in which we examined the relationship between the mean squared error (MSE) and 
the number of bins $N$.
We first generated 
a set of sparse signals $\bm s_1, \bm s_2,\ldots, \bm s_{M}$
according to the Bernoulli-Gaussian distribution with 
$p = 0.1$ and non-zero elements following ${\cal N}(0,1)$.
The MSE at time $t$ is defined by ${\sf E}[\|\bm x(t) - \bm s \|_2^2]$,
which can be estimated by 
\begin{align}
	{\sf MSE}(t) \equiv {\sf E}[\|\bm x(t) - \bm s \|_2^2] 
	\simeq \frac{1}{M} \sum_{i \in [M]}\|\bm x^{(k)}_i - \bm s_i\|_2^2,
\end{align}
where $\bm x^{(k)}_i$ is the output of the Euler method for the $i$th trial at index $k$
corresponding to time $t$.
The parameter settings are nearly the same as the settings in the previous experiment; that is,
$n=128, m = 64, \sigma = 0.1, \lambda = 5, \alpha = 50$. The number of samples was set to 
$M = 1000$.

Table \ref{N_dependency} shows the estimates of MSE for several different discretizations.
Parameter $T$ for the discretizations  
was set to 4. From Table \ref{N_dependency}, we can see that 
the values of the MSE estimates converge as $N$ increases.
This result provides evidence that the Euler method works adequately for this ODE.
In the following experiments, we use $N = 5000$ to ensure 
good accuracy and a reasonable computational time.

\begin{table}[htbp]
\centering
\caption{Number of bins $N$ and estimated MSE ($T=4$)}
\label{N_dependency}
\begin{tabular}{cc}
\hline
\hline
$N$ & ${\sf MSE}(T)$ \\
\hline
1000 & 0.503796 \\
2000 & 0.393782 \\
5000 & 0.408937 \\
10000 &0.408701 \\
\hline 
\end{tabular}
\end{table}

\begin{figure}[htbp]
\begin{center}
\includegraphics[width=\columnwidth]{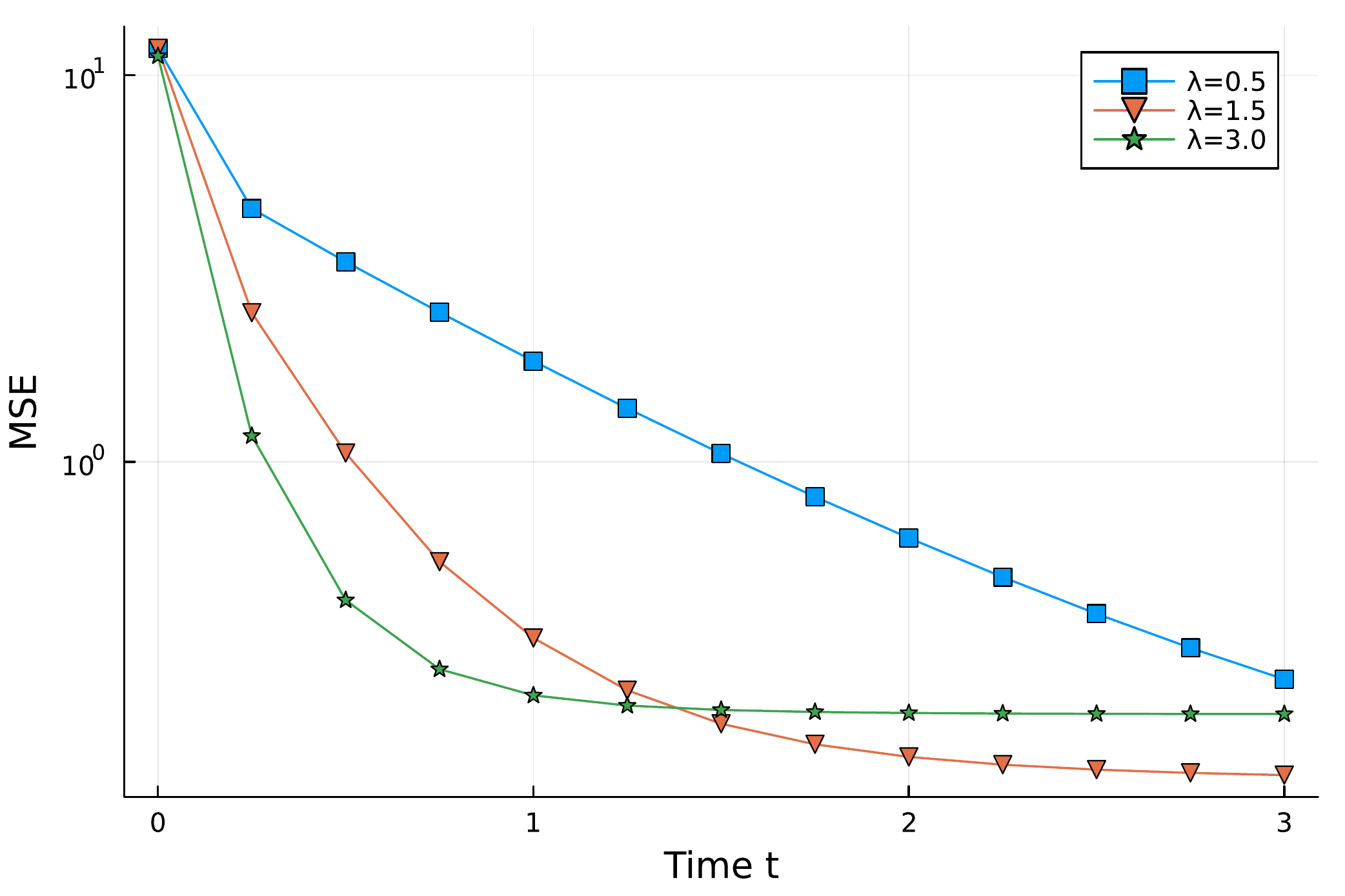}
\caption{Estimates of mean squared error ${\sf MSE}(t)$ as a function of time $t$ ($n=128, m = 64, p=0.1, \sigma=0.1, \alpha=50$).}
\label{ODE-ISTA_MSE}
\end{center}
\end{figure}

Figure \ref{ODE-ISTA_MSE} plots the estimates of ${\sf MSE}(t)$ as a function 
of time $t$.
The MSE estimates in Fig.~\ref{ODE-ISTA_MSE}
are based on
100 trials of sparse signal recovery processes, i.e., $M = 100$.
The Euler method with $N = 5000$ was used for $T = 4$ to estimate 
${\sf MSE}(t) (0 \le t \le T)$.
The parameter settings are nearly the same as in the previous experiment.
We examined three cases: $\lambda = 0.5, 1.5, 3$.
From Fig.~\ref{ODE-ISTA_MSE}, it is apparent that  
the MSE curve for $\lambda = 0.5$ has the largest MSE 
in the range $0 \le t \le 3$; however, the MSE value continues to decrease 
after $t = 3$. On the other hand, the MSE curve for $\lambda = 3$
shows the fastest convergence, but the value of ${\sf MSE}(t)$ saturates 
to a relatively high constant value in the range $t > 1.0$.
The MSE curve for $\lambda = 1.5$ provides a slightly slower convergence 
compared with the case of $\lambda = 3$, but it shows a lower 
floor for the MSE values. 
These experimental results imply that the regularization constant $\lambda$ 
has a strong influence on convergence behavior, similar to discrete time 
sparse signal recovery algorithms such as ISTA.

\section{Local Convergence Analysis}
\label{sec:localconveregnce}

\subsection{Linear approximation around equilibrium point}

In the previous section, 
we saw that the convergence behavior of the ODE (\ref{diff_eq}) is highly dependent
on the choice of the regularization parameter $\lambda$. 
In this section, the mechanism by which the regularization parameter affects the convergence behavior 
is examined. The core of the analysis presented here is the use of 
a linear approximation around the equilibrium point~\cite{Strogatz}  of the ODE 
to study the local convergence behavior.

In the following analysis, we consider 
the objective function (\ref{def_g}).
The gradient of $g$ is thus given by
\begin{align}
	\nabla g(\bm x) = \bm A^T (\bm A \bm x - \bm y) + \lambda \tanh(\alpha \bm x).
\end{align}
We can now examine the local convergence behavior of the continuous-time dynamical system defined by
\begin{align} \label{general_ode}
	\frac{d \bm x(t)}{dt} &= 
	-\left(\bm A^T (\bm A \bm x(t) - \bm y) + \lambda \tanh(\alpha \bm x(t) ) \right), \\
	\bm x(0) &= \bm x_0,
\end{align}
where the initial vector $\bm x(0) = \bm x_0$ is assumed to be given as
a boundary condition.

The equilibrium point $\bm x^*$ of the above ODE 
is the point satisfying $\nabla g(\bm x) = \bm 0$.
Thus, the equilibrium point $\bm x^*$ satisfies the equality
\begin{align} \label{equilibrium_condition}
	\bm A^T (\bm A \bm x^* - \bm y) + \lambda \tanh(\alpha \bm x^*) = \bm 0,
\end{align}
which is called {\em equilibrium equality}.

To establish the dynamics of the residual error, 
we change the coordinate of the ODE according to 
\begin{align}
	\bm x(t) = \bm x^* + \bm e(t),	
\end{align}
where $\bm e(t) \equiv \bm x(t) - \bm x^*$ represents the residual error vector. 
In the following analysis, $\bm e(t)$ is abbreviated to $\bm e$ for simplicity.
Due to the change in the coordinate, 
the right-hand side of ODE (\ref{general_ode})
can be transformed to
\begin{align}
	-&\left(\bm A^T (\bm A \bm x - \bm y) + \lambda \tanh(\alpha\bm x)\right) \\
	&=
		-\left(\bm A^T (\bm A (\bm x^* + \bm e) - \bm y) + \lambda \tanh(\alpha(\bm x^* + \bm e)) \right) \\
		&=-\left(\bm A^T (\bm A \bm x^* - \bm y) + \bm A^T \bm A\bm e + \lambda \tanh(\alpha(\bm x^* + \bm e))\right) \\ \label{eq_tmp}
		&= - \left(\bm A^T \bm A \bm e + \lambda \tanh(\alpha(\bm x^* + \bm e)) -  \lambda \tanh(\alpha\bm x^*)\right).
\end{align}
The last equality is due to the equilibrium equality.

By expanding $\tanh(\alpha \bm x)$ using the Taylor expansion 
and ignoring the higher-order terms, we can obtain the 
linear approximation around the equilibrium point:
\begin{align}
	\tanh(\alpha\bm x) \simeq \tanh(\alpha\bm x^*) + \bm J(\bm x^*) (\bm x - \bm x^*),
\end{align}
where $\bm J(\bm x^*)$ is the Jacobian matrix of $\tanh(\alpha \bm x)$ at $\bm x^*$.
The Jacobian matrix is given by
\begin{align}
J(\bm x^*)
= \mbox{diag}\left(\frac{\alpha}{\cosh^2(\alpha x_1^*)},\frac{\alpha}{\cosh^2(\alpha x_2^*)},\ldots, \frac{\alpha}{\cosh^2(\alpha x_n^*)} \right),
\end{align}
where $\bm x^* = (x_1^*, x_2^*,\ldots, x_n^*)^T$.
Note that $\cosh^2(x) > 0$ for any $x \in \mathbb R$. This implies that 
$J(\bm x^*)$ is positive definite for any $\bm x^* \in \mathbb{R}^n$.

By using the linear approximation, we immediately have 
the approximation:
\begin{align}
	 \lambda \tanh(\alpha(\bm x^* + \bm e)) -  \lambda \tanh(\alpha \bm x^*)
	 &\simeq \lambda J(\bm x^*) \bm e
\end{align}
In the following argument, we will treat this approximated equality as an equality for the sake of simplicity.
Substituting this equation into (\ref{eq_tmp}), we obtain the
linear approximation of the right-hand side of (\ref{general_ode}):
\begin{align}
	-\left(\bm A^T (\bm A \bm x - \bm y) + \lambda \tanh(\alpha \bm x) \right) &=
		 - \left(\bm A^T \bm A +  \lambda \bm J(\bm x^* ) \right)  \bm e 
\end{align}
Finally, we have the ODE representing the evolution of the 
residual error vector: 
\begin{align}
	\frac{d \bm e(t)}{dt} &=- \left(\bm A^T \bm A +  \lambda \bm J(\bm x^* ) \right)  \bm e(t), \\
	\bm e(0) &= \bm e_0,
\end{align}
where $\bm e_0$ represents the initial error vector.
This is a linear ODE, which can be solved in 
a concise form \cite{NumericalMethods}:
\begin{align} \label{err_solution}
	\bm e(t) = \exp \left( - \left(\bm A^T \bm A +  \lambda \bm J(\bm x^* )\right)  t  \right) \bm e(0).
\end{align}
We can expect the above solution to reflect the actual error behavior well 
if the initial error $\bm e(0)$ is sufficiently close to the zero vector, 
i.e., in the case where the linear approximation is appropriate.

From the error evolution equation (\ref{err_solution}), it is clear that 
the eigenvalues of the matrix $\left(\bm A^T \bm A +  \lambda \bm J(\bm x^* )\right)  t$ dominate
the behavior of the error evolution. Due to the positive definiteness of 
the Jacobian matrix $\bm J(\bm x^*)$
and the semi-positive definiteness of the Gram matrix $\bm A^T \bm A$,
all the eigenvalues of the matrix 
$\left(\bm A^T \bm A +  \lambda \bm J(\bm x^* )\right) t$ are positive real numbers.
This means that the time evolution of error vector (\ref{err_solution}) is 
{\em locally asymptotically stable}; that is,
$
	\lim_{t \rightarrow \infty} \bm e(t) = \bm 0	
$
for any initial error $\bm e(0)$ sufficiently close to the zero vector.

\subsection{Upper bound on convergence speed}

From the time evolution or error vector (\ref{err_solution}),
we immediately have the following theorem, under the assumption that the linear approximation is appropriate.
\begin{theorem} \label{main_th}
If the initial vector $\bm x(0)$ is sufficiently close to $\bm x^*$, 
then the following inequality holds:
\begin{align}
	 \frac{\|\bm e(t) \|_2}{\|\bm e(0)\|_2} 
	&\le \exp(-\omega_1 t),
\end{align}
where $\{\omega_i \}_{i=1}^n$ is the set of eigenvalues of 
$
\bm A^T \bm A +  \lambda \bm J(\bm x^* ),
$
with the order $0 < \omega_1 \le \omega_2 \le \cdots \le \omega_n$.
\end{theorem}

(Proof)
Taking the norm of both sides of (\ref{err_solution}), we have
an inequality:
\begin{align}
	\|\bm e(t) \|_2 
	&= \|\exp \left( - \left(\bm A^T \bm A +  \lambda \bm J(\bm x^* )\right)  t  \right) \bm e(0)\|_2 \\
	&\le \|\exp \left( - \left(\bm A^T \bm A +  \lambda \bm J(\bm x^* )\right)  t  \right)\|_2 \| \bm e(0)\|_2,
\end{align}
where the matrix norm is the spectral norm.
By dividing both sides by $\| \bm e(0)\|_2$, we obtain
\begin{align} \label{fund_ineq}
	\frac{\|\bm e(t) \|_2}{\|\bm e(0)\|_2} 
	&\le \|\exp \left( - \left(\bm A^T \bm A +  \lambda \bm J(\bm x^* )\right)  t  \right)\|_2. 
\end{align}
Note that the above matrix exponential can be rewritten as
\begin{align}
\exp \left( - \left(\bm A^T \bm A +  \lambda \bm J(\bm x^* )\right)  t  \right) 
= \bm U \mbox{diag}(e^{-\omega_1 t}, \ldots,e^{-\omega_n t})\bm U^{T},
\end{align}
where $\bm U$ is an orthogonal matrix.
Since the spectral norm of a semi-positive definite symmetric matrix $\bm X$ coincides with the  
largest eigenvalue of $\bm X$, we have
\begin{align}
\|\exp \left( - \left(\bm A^T \bm A +  \lambda \bm J(\bm x^* )\right)  t  \right)\|_2= \exp(-\omega_1 t).
\end{align}
Substituting this equality into $(\ref{fund_ineq})$, we obtain the claim of the theorem. \hfill \fbox{}

The above theorem indicates that the term $\exp(-\omega_1 t)$ dominates 
the convergence speed around an equilibrium point.
The average behavior is described by the following corollary, 
which can be derived directly 
from Theorem \ref{main_th}.
\begin{corollary}
If the initial vector $\bm x(0)$ is sufficiently close to $\bm x^*$, 
then the following inequality holds:
\begin{align} \label{col}
	{\sf E} \left[\frac{\|\bm e(t) \|_2}{\|\bm e(0)\|_2}  \right]
	&\le {\sf E}[\exp(-\omega_1 t)].
\end{align}
\end{corollary}

Let $l_{min}(\bm X)$ represent the minimum eigenvalue of a symmetric matrix $\bm X$.
The minimum eigenvalue of $\bm A^T \bm A +\lambda \bm J(\bm x^*)$ satisfies
\begin{align}
	l_{\sf min}(\bm A^T \bm A +\lambda \bm J(\bm x^* )) 
	&\ge l_{\sf min}(\bm A^T \bm A) +  l_{\sf min}(\lambda \bm J(\bm x^* )) \\
	& =  l_{\sf min}(\bm A^T \bm A) + \lambda \min_{i \in [n]} \frac{\alpha}{\cosh^2(\alpha x_i^*)}.
\end{align}
Therefore, increasing the value of $\lambda$ leads to larger $\omega_1$,
which increases the convergence speed. 

\subsection{MSE behavior}
\label{mse_behavior}

The value on the right-hand side of equation (\ref{col}) can be 
reduced by choosing a larger $\lambda$. This implies that 
a larger $\lambda$ will achieve faster convergence to the equilibrium point. 
This qualitative argument explains the numerical results shown in Fig.~\ref{ODE-ISTA_MSE}; 
in particular, a larger $\lambda$ results in a faster decrease of ${\sf MSE}(t)$.

On the other hand, from the equilibrium equation (\ref{equilibrium_condition}), we have
\begin{align}
	\|\bm A^T (\bm A \bm x^* - \bm y)\|_2  =  \lambda \| \tanh(\alpha \bm x^*)\|_2.
\end{align}
Consider a scenario where the noise is negligible.
Recall that the original sparse vector $\bm s$ should satisfy
\begin{align}
	\|\bm A^T (\bm A \bm s - \bm y)\|_2  \simeq 0
\end{align}
under the assumption of negligible noise.
In such a case, an increase in $\lambda$ can lead to an increase in the discrepancy $\|\bm x^* - \bm s \|_2$.
From this observation, it can be concluded that a larger $\lambda$ leads to higher MSE floors,
which is indeed observed in Fig.~\ref{ODE-ISTA_MSE}.

\subsection{Numerical experiments}

The analysis presented in the previous sections is 
based on a linear approximation around the equilibrium point. 
Therefore, the accuracy of this approximation should be verified by numerical experiments.

The error norm ratio $\rho(t)$ is defined by
\begin{align}
\rho(t) \equiv \frac{\|\bm e(t) \|_2}{\|\bm e(0)\|_2}  = \frac{\|\bm x(t) -  \bm x^*\|_2}{\|\bm x(0) - \bm x^*\|_2}.
\end{align}
In the following experiment, we will evaluate the norm ratio $\rho(t)$ for the two cases 
$\lambda = 1.5$ and $\lambda = 5$.

The parameter settings are nearly the same as in the previous experiment:
$n=128, m = 64, \sigma = 0.1,\alpha = 50$. 
The equilibrium point $\bm x^*$ is approximated by the value of $\bm x(T)$ where $T=4$.

We consider two types of initial points, $\bm x(0) = \bm 0$ and $\bm x(0)
= \hat{\bm x} \equiv \bm x^* + \bm \epsilon$,
where $\bm \epsilon \sim {\cal N}(\bm 0, 0.1^2 \bm I)$.
Note that $\bm x(0) = \bm 0$ may not be in the region where the linear approximation is valid but
$\hat{\bm x}$ can be in such a region because the norm of $\bm \epsilon$ is so small.

\begin{figure}[htbp]
\begin{center}
\includegraphics[width=\columnwidth]{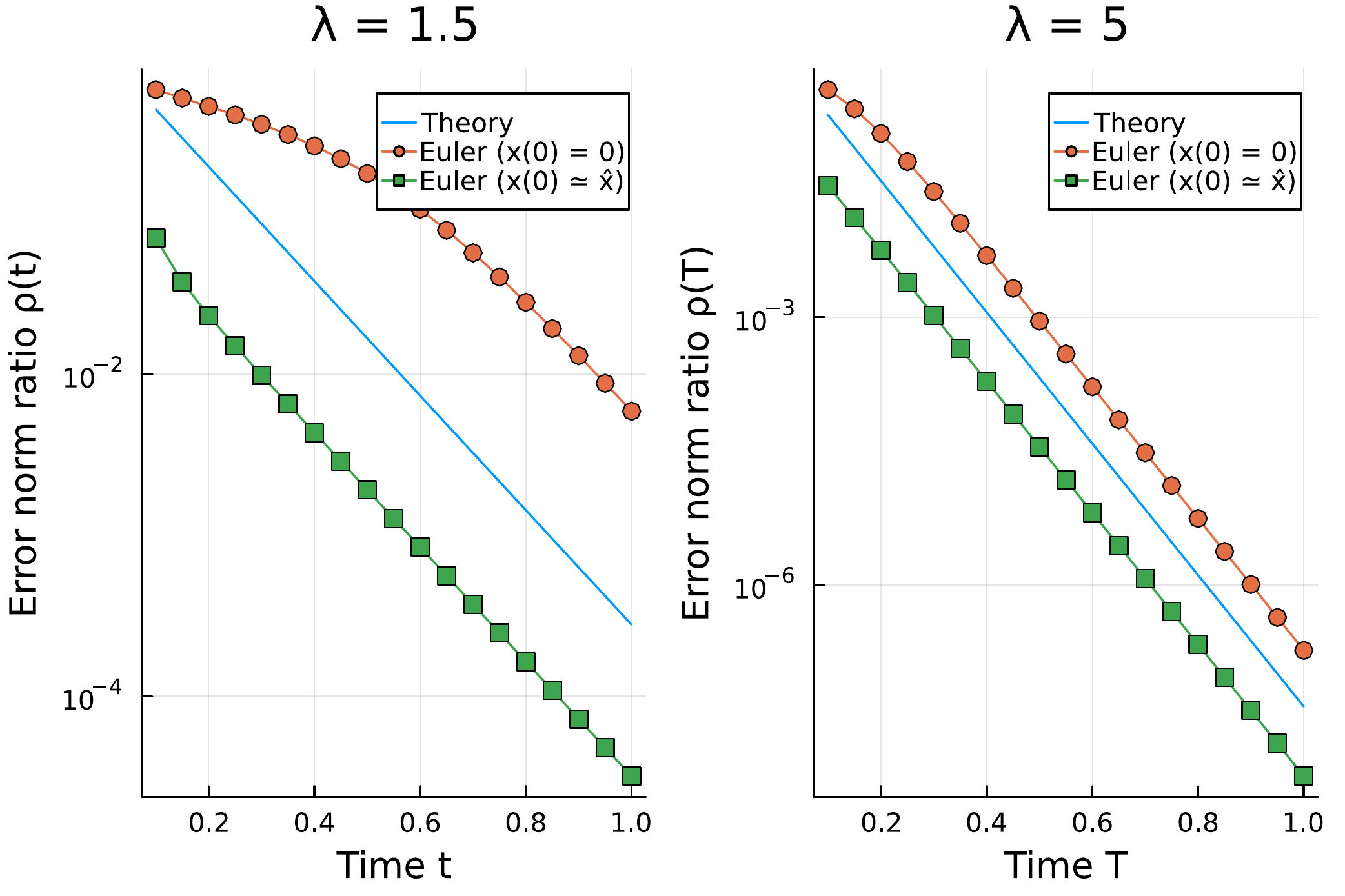}
\caption{Error norm ratio $\rho(t)$ as a function of time $t$. Left $\lambda = 1.5$, Right $\lambda = 5$ ($n=128, m = 64, p=0.1, \sigma=0.1,\alpha=50$).}
\label{convergence}
\end{center}
\end{figure}

Figure \ref{convergence} displays the error norm ratio $\rho(t)$ as a function of time $t$. 
The solid line labeled ``Theory'' corresponds to the value of $\exp(-\omega_1 t)$. 
The left panel shows the case where $\lambda = 1.5$. 
The error norm ratio curve of $\rho(t)$ with the initial condition $\bm x(0) = \bm 0$ 
has a shallower slope than that of the curve of $\exp(-\omega_1 t)$ when $t < 0.7$. 
This is due to the invalidity of the linear approximation for small $t$. 
It can be observed that the curve of $\rho(t)$ with $\bm x_0 = \bm 0$ 
has almost the same slope as that of $\exp(-\omega_1 t)$ when $t > 0.7$. 
This can be explained by the fact that 
the state point $\bm x(t)$ gradually enters the region where the linear approximation is valid.
Once the state point enters such a region, the linear approximation theory can be applied.
On the other hand, in the case of $\bm x(0) = \hat{\bm x}$, 
the slope of $\rho(t)$ is almost the same as that of $\exp(-\omega_1 t)$ from the beginning. 
The result is consistent 
with the claim  of Theorem \ref{main_th}. 
Since $\hat{\bm x}$ is sufficiently close to $\bm x^*$, 
the linear approximation is valid from the start.

The right panel shows the numerical results for $\lambda = 5$. 
Nearly the same observations can be made as in the left panel. Compared to the left panel, 
the value of $\rho(t)$ is much smaller for each $t$. 
This observation is consistent with the argument concerning the relationship 
between $\lambda$ and the convergence rate in the previous section.
 
Note that in all the cases, 
the asymptotic slope of $\rho(t)$ can be accurately estimated by the slope of $\exp(-\omega_1 t)$. 
This observation supports our claim that the smallest eigenvalue $\omega_1$ 
determines the convergence rate around an equilibrium point. 
The numerical results presented in Fig.~\ref{convergence} can be seen
as a numerical validation of the argument in the previous sections.

\begin{figure}[htbp]
\begin{center}
\includegraphics[width=\columnwidth]{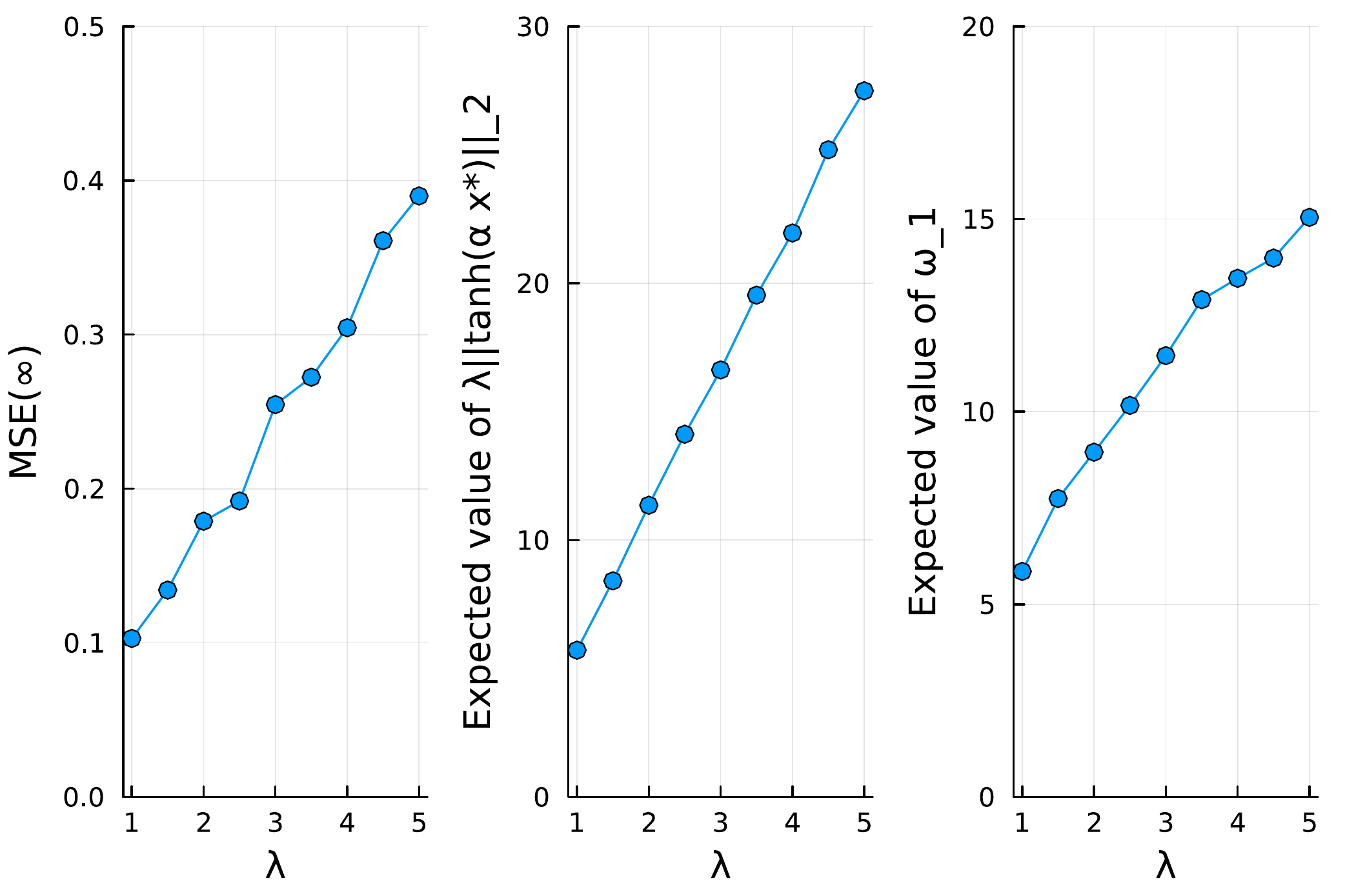}
\caption{Values of ${\sf MSE}(\infty)$, 
${\sf E}[|\lambda \tanh(\alpha \bm x^*)\|_2^2]$, and ${\sf E}[\omega_1]$
from left to right ($n=128, m = 64, p=0.1, \sigma=0.1, \alpha=50$).}
\label{lambda_vs_x}
\end{center}
\end{figure}

Figure \ref{lambda_vs_x} shows 
${\sf MSE}(\infty)$, 
${\sf E}[|\lambda \tanh(\alpha \bm x^*)\|_2^2]$, and ${\sf E}[\omega_1]$
estimated from 500 trials. For estimating the asymptotic MSE
${\sf MSE}(\infty)$, we used $T = 4$ rather than $T = \infty$. The numerical results of
${\sf MSE}(\infty)$ indicate that the MSE floor increases as $\lambda$
increases. We can also observe that the values 
${\sf E}[|\lambda \tanh(\alpha \bm x^*)\|_2^2]$ (center panel) 
is an increasing function of $\lambda$.
This observation is consistent with the argument in 
Subsection \ref{mse_behavior}.
The local convergence rate is determined by $\omega_1$.
The tendency of ${\sf E}[\omega_1]$ is shown 
in the third panel of Fig.~\ref{lambda_vs_x}.
It can be confirmed that a larger $\lambda$ achieves faster convergence.

\begin{figure}[htbp]
\begin{center}
\includegraphics[width=\columnwidth]{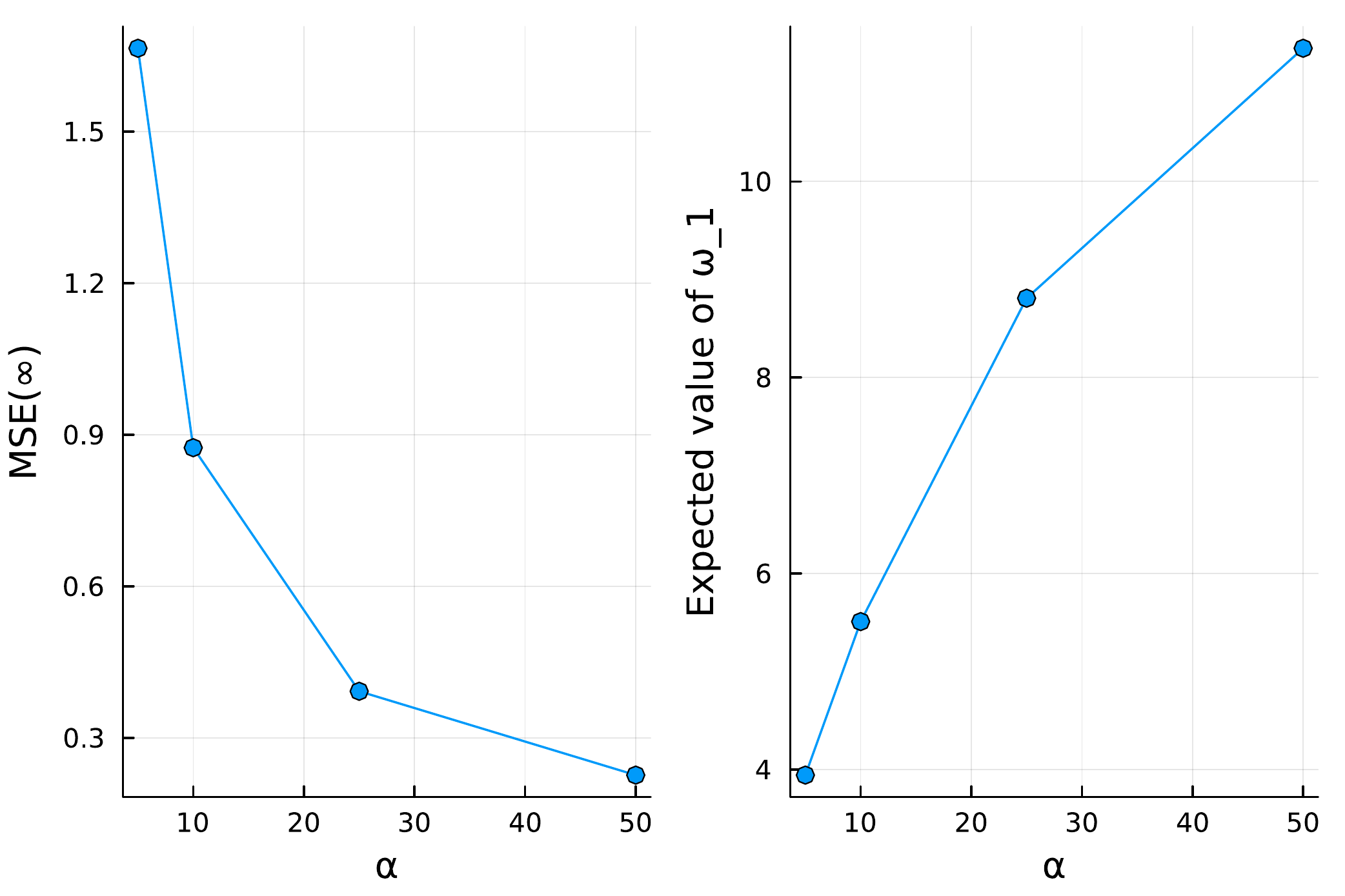}
\caption{Values of ${\sf MSE}(\infty)$ and ${\sf E}[\omega_1]$
($n=128, m = 64, p=0.1, \sigma=0.1, \lambda = 3$).}
\label{lambda_vs_alpha}
\end{center}
\end{figure}

Finally, we will experimentally confirm the relationship between $\alpha$ 
and the MSE convergence behavior.
Figure \ref{lambda_vs_alpha} shows the values of ${\sf MSE}(\infty)$ and ${\sf E}[\omega_1]$
where the value of $\lambda$ is fixed to 3. From the results of ${\sf MSE}(\infty)$, 
we can say that a larger $\alpha$ results in a smaller MSE floor. 
The right panel showing ${\sf E}[\omega _1]$ indicates that a larger $\alpha$ leads to 
faster convergence. In summary, $\alpha$ should be large in order to achieve
faster convergence and a high-quality solution. 
We therefore use a relatively large constant $\alpha = 50$ throughout this paper.

\section{Deep Unfolded-Variational  Optimization}
\label{sec:DU}
\subsection{Parametric ODE and its optimization}
\label{homotopy}

\begin{figure}[htbp]
\begin{center}
\includegraphics[width=\columnwidth]{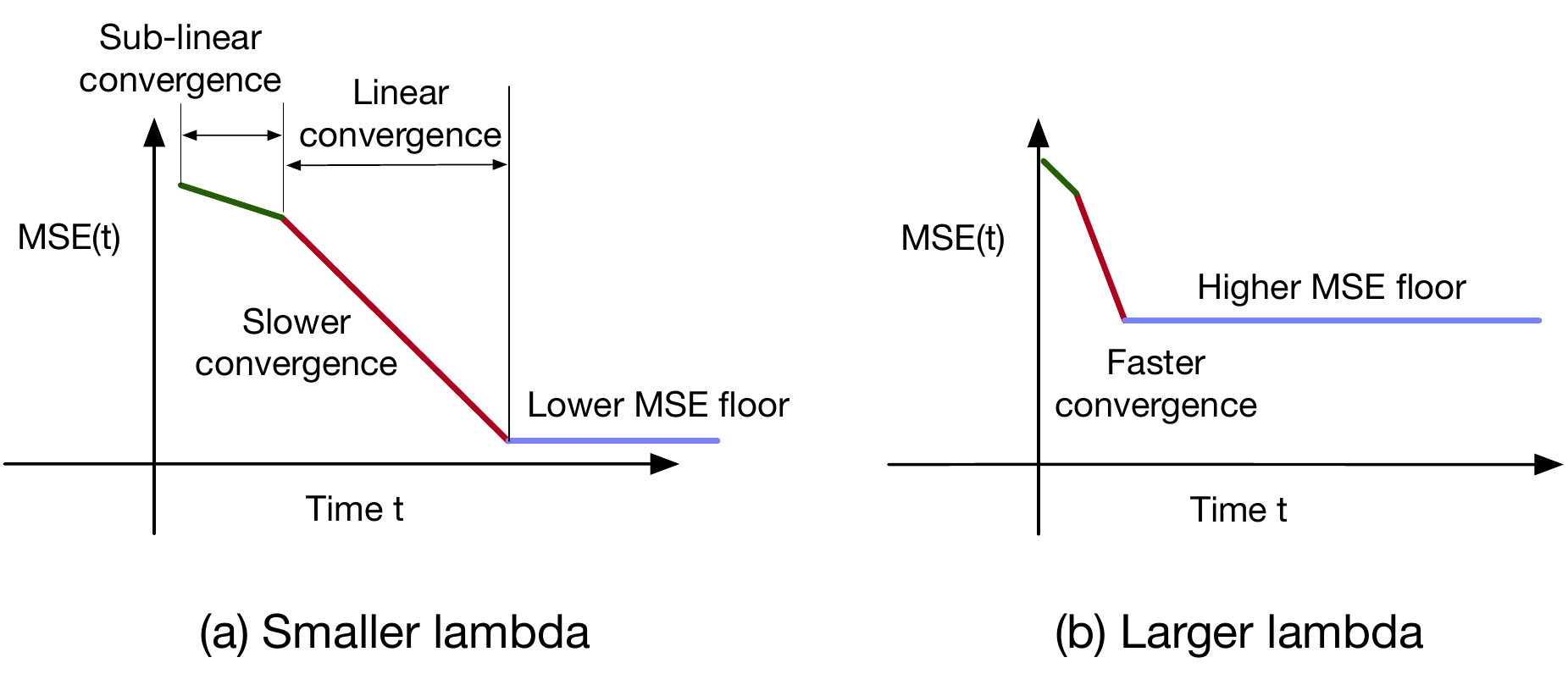}
\caption{Typical tendency of the convergence behavior of ${\sf MSE}(t)$}
\label{lambda_effect}
\end{center}
\end{figure}

A summary of the theoretical analysis and numerical results presented 
in the previous section is illustrated in Fig.~\ref{lambda_effect}, 
which shows the typical behavior of the convergence of ${\sf MSE}(t) = {\sf E}[\|\bm x(t) - \bm s \|_2^2]$.
As discussed earlier, there is a trade-off between the MSE floor value 
(the asymptotic MSE, ${\sf MSE}(\infty)$) and the convergence rate.
If we use a large $\lambda$, we can expect fast convergence, but
we will need to compromise on the quality of the solution. On the other hand,
if we use a small $\lambda$, we must allow for slow convergence.

An important consideration is the time period during which sub-linear convergence occurs. 
If the state point $\bm x(t)$ is close enough to the equilibrium point, 
the linear approximation can be applied and linear convergence can be expected, 
as discussed in a previous section. However, if $\bm x(t)$ is not close enough to the equilibrium point, 
linear convergence cannot be expected. The left panel of Fig.~\ref{convergence} 
illustrates such a regime, specifically for $0 \le t \le 0.6$. 
As shown in Fig.~\ref{lambda_effect}, the linear convergence regime appears after a sub-linear regime.

To improve both convergence speed and solution quality, 
one strategy is to introduce a time-dependent regularization parameter, i.e., 
varying the value of $\lambda$. Starting with a large $\lambda$ shortens the sub-linear regime
and increases the convergence rate. 
Changing $\lambda$ judiciously keeps $\bm x(t)$ in the linear convergence regime. 
By adjusting $\lambda$ to take on the smallest value at the end of the process, 
a high-quality solution can be obtained.

\begin{figure}[htbp]
\begin{center}
\includegraphics[scale=0.55]{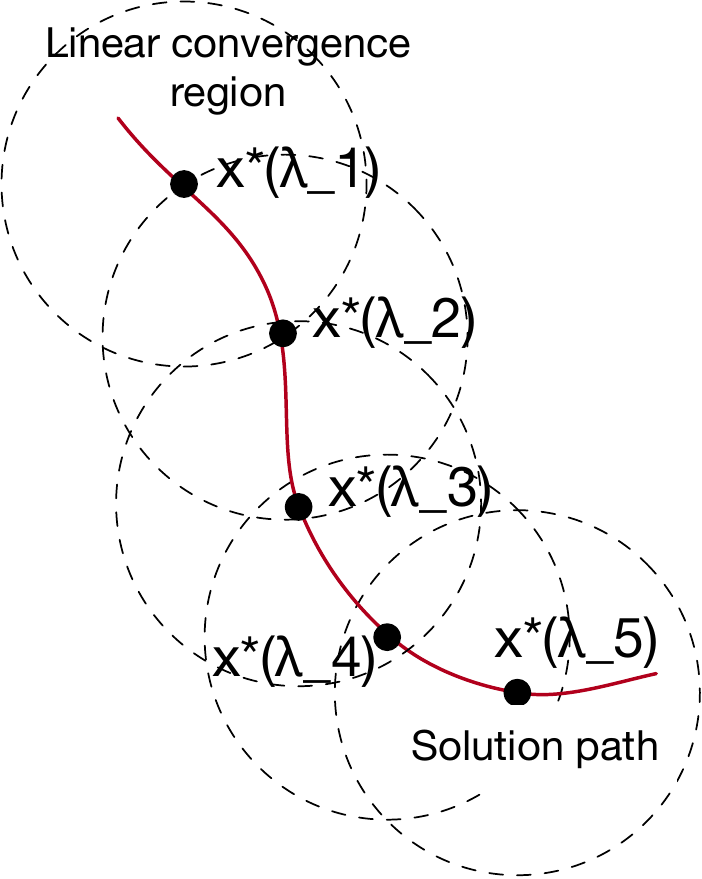}
\caption{Concept of continuation method: $x^*(\lambda_i)(i \in [5])$ represents the equilibrium point when $\lambda = \lambda_i$. Using an appropriately chosen set of parameters $\{\lambda_i\}$, the solution path is fully covered by the linear convergence regions. We can anticipate both faster convergence and a 
	high-quality solution in such a case.
 }
\label{continuation}
\end{center}
\end{figure}

This idea is closely related to the {\em homotopy or continuation method}, 
where the value of the regularization parameter is modulated 
so that the state vector progresses along the solution path. 
For example, Xiao and Zhang \cite{Xiao} proposed a technique for solving the Lasso problem 
by using a proximal algorithm, which incorporates a sequence 
of decreasing values of the regularization parameter. 
The concept of the continuation method is shown in Fig.~\ref{continuation}.

The above idea naturally leads to the following {\em parametric ODE}:
\begin{align} \label{parametric_eq}
	\frac{d \bm x(t)}{dt} 
	&= - \left(\bm A^T (\bm A \bm x(t) - \bm y) + \lambda(t) \tanh(\alpha \bm x(t)) \right),\\
	\bm x(0) &= \bm x_0,
\end{align}
where the regularizing constant is replaced with the
function $\lambda: \mathbb{R} \rightarrow \mathbb{R}$.

In this section, we examine the optimization 
of $\lambda(t)$ in this parametric ODE. The optimization problem at hand can be formulated as
\begin{align} \label{opt_prob}
\mbox{minimize } {\sf E}[\|\bm x(T^*) - \bm s \|_2^2] \ 
\mbox{subject to } \lambda: \mathbb{R} \rightarrow \mathbb{R},
\end{align}
where $T^*$ is the predetermined target time. 
In essence, we aim to find the optimal continuous-time schedule for $\lambda$ 
that produces a high-quality solution at a given time $T^*$. 
This optimization problem falls into the category of variational optimization problems. 
Unfortunately, we cannot expect to derive a compact analytical solution to this problem, 
and, thus, it must be solved numerically.

\subsection{Deep unfolded-variational optimization (DU-VO) method}

The development of deep neural networks has also 
had a significant impact on the design of 
algorithms for communications and signal processing~\cite{Com1,Com2,Com3}.
{\em Deep unfolding}, as described in works such as \cite{LISTA, DU, Ito}, 
can be seen as a highly effective method for improving the convergence of iterative algorithms.
Gregor and LeCun introduced the Learned ISTA (LISTA) \cite{LISTA}, which uses learnable matrices.
LISTA achieves a recovery performance that is much
superior to that of the original ISTA. Borgerding et al. also presented variants of AMP and VAMP 
with learnable capability~\cite{LAMP}\cite{Borgerding}.
Trainable ISTA (TISTA)~\cite{Ito} is a recent learnable sparse signal recovery algorithm 
with fast convergence.
TISTA requires a small number of trainable parameters, which provides a fast and stable training process.  

The concept of deep unfolding is simple: 
Embed trainable parameters in the original iterative algorithm, 
followed by the unfolding of the signal-flow graph of the original algorithm. 
The standard supervised training techniques used in deep learning, such as 
stochastic gradient descent (SGD)~\cite{SGD} and back propagation~\cite{BackProp}, 
can then be applied to the unfolded signal-flow graph to optimize the trainable parameters.

In the following, we will introduce the {\em deep unfolded-variational optimization} (DU-VO) method 
for the numerical solution of variational optimization problems. 
The combination of deep unfolding and the Euler method for differential equation solvers~\cite{Rackauckas2} 
is a current area of active research in scientific machine learning. 
However, it should be noted that the technique is not limited to applications within scientific machine learning; 
rather, it seems to be a powerful tool for 
{\em optimizing continuous-time dynamical systems}.

This subsection deals with variational optimization problems in a general context. 
We will return to the specific optimization problem defined in (\ref{opt_prob}) later.

Assume that we have an ODE
\begin{align} \label{xevolution}
	\frac{d\bm x(t)}{dt} = h(\bm x(t), u(t), t ) 
\end{align}
and it defines the dynamics of $\bm x(t)$.
For a given integrable function $F$,  
a functional $J$ to be minimized is defined by 
\begin{align} \label{ufunctional}
	J[u(t)] \equiv \int_{0}^T F(\bm x(t) , u(t), t) dt.
\end{align}
In the following argument, we assume the case where $u(t)$ is one-dimensional, i.e., 
$u: \mathbb{R} \rightarrow \mathbb{R}$, and can control this function.
We here consider a variational optimization problem including $\bm x(t)$ 
defined by an ODE (\ref{xevolution}):
\begin{align} \label{vo_prob}
	\mbox{minimize }_{u(t)}\ J[u(t)].
\end{align}
Namely, we want to find a one-dimensional function $u$ minimizing the value of the functional $J$.

This type of variational optimization problem often appears 
in the field of optimal control \cite{StochasticHJB}. 
The optimal solution of (\ref{vo_prob}) can be obtained by solving a Hamilton-Jacobi-Bellman (HJB) equation \cite{StochasticHJB}. 
However, solving the HJB equation is 
generally difficult because it requires solving a non-linear partial differential equation. 
In a stochastic context, solving a stochastic HJB equation \cite{StochasticHJB} 
becomes even more complex and challenging.

The DU-VO method is a numerical approach 
that aims to solve the variational problem (\ref{vo_prob}). 
The first step in the implementation of the DU-VO method is to introduce 
a function approximation using a radial basis function (RBF), which is used to approximate $u(t)$.

\begin{figure}[htbp]
\begin{center}
\includegraphics[height=5cm]{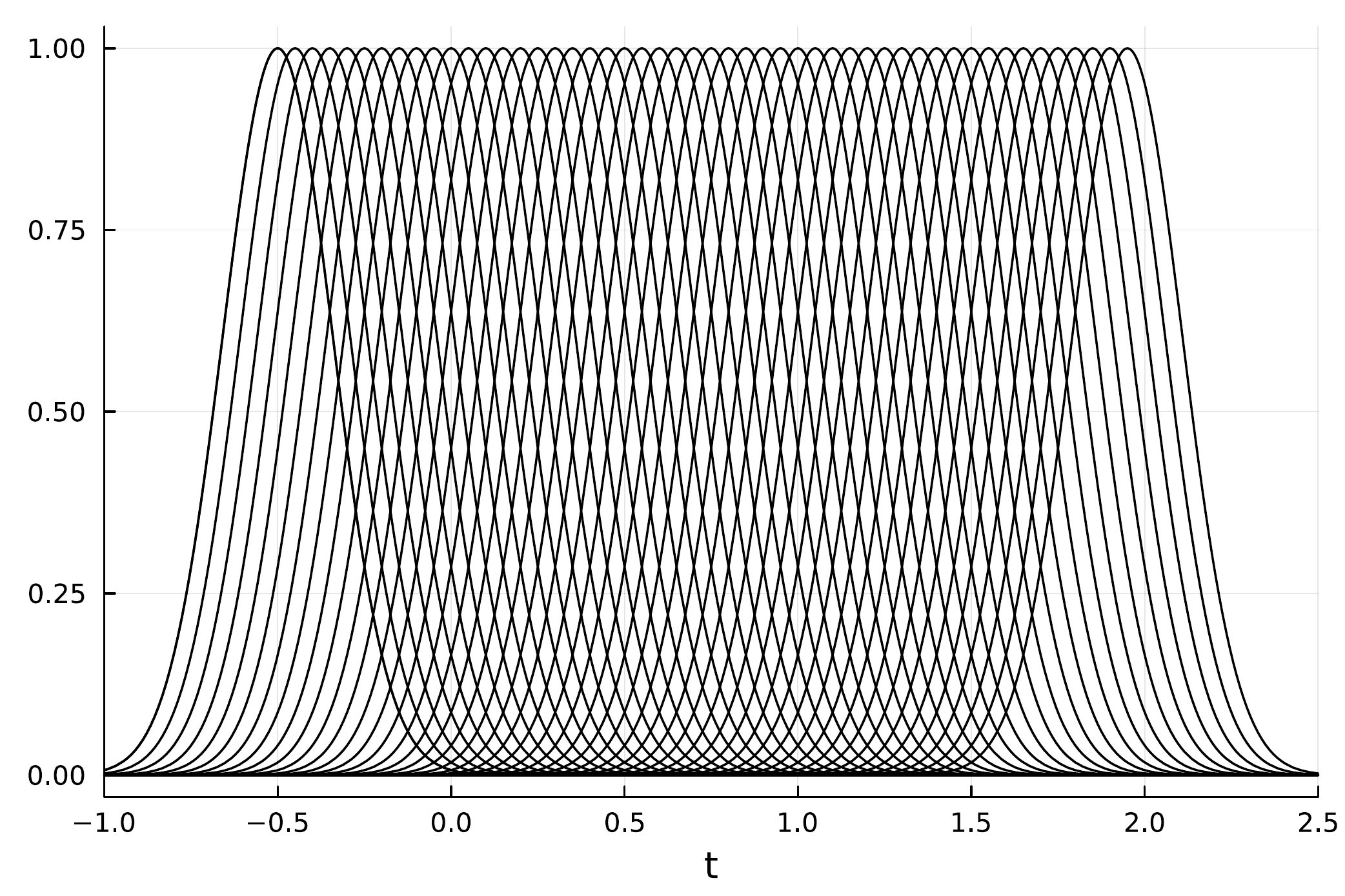}
\caption{Set of Gaussian RBFs: $\phi(t - c_i) = \exp( -\beta(t-c_i)^2)$. $\beta = 20$, 
$c_1 = -0.5, c_2 = -0.45,\ldots, c_m = 2$.}
\label{fig:rbf_shape}
\end{center}
\end{figure}

Let $\phi: \mathbb{R} \rightarrow \mathbb{R}$ be an RBF,
which satisfies
$
\phi(x) = \phi(|x|).
$
A continuous function $u(t)$ can be approximated by 
	\begin{align} \label{rbf_aprox}
		\tilde{u}(t) \equiv \sum_{i=1}^S w_i \phi(t - c_i),
	\end{align}
where the weight vector 
$$
\bm w \equiv (w_1,w_2,\ldots, w_S)^T \in \mathbb{R}^S
$$ 
is a trainable parameter that can be updated in an optimization process.
The above function approximation is known as the {\em RBF approximation}.
The shift parameter 
\begin{align}
\bm c \equiv (c_1, c_2,\ldots, c_S)^T \in \mathbb{R}^S	
\end{align}
is treated as a 
hyperparameter.
Thus, we will use the parametric model of $u(t)$ defined by (\ref{rbf_aprox})
in the following discussion.
Figure \ref{fig:rbf_shape} shows a set of Gaussian RBFs defined by
$
\phi(t - c_i) = \exp( -\beta(t-c_i)^2).
$
We use the Gaussian RBF in the numerical computation below.

The advantage of using the RBF is that it requires only a small number of trainable parameters 
to approximate a one-dimensional continuous function, 
resulting in a stable and efficient optimization process.

As in Subsection \ref{Euler_sec}, 
the closed interval $\{t \mid 0 \le t \le T \}$ is divided into $N$ bins, and 
the discrete time indices are defined as $t^{(k)} \equiv \eta k (k =0,1,\ldots, N)$,
where $\eta \equiv T/N$ is the width of the bins.
We also introduce the discretized state vector
$\bm x^{(k)}\in \mathbb{R}^n$ 
corresponding to $\bm x(t^{(k)})$.

\begin{algorithm}
 \caption{DU-VO method}
 \label{opt_alg}
 \begin{algorithmic}[1]
 \renewcommand{\algorithmicrequire}{\textbf{Input:}}
 \renewcommand{\algorithmicensure}{\textbf{Output:}}
  \REQUIRE $\bm x_0$ (initial state), $\bm w_0$ (initial weight)
  \ENSURE $\bm w$ (optimized weight)
  \STATE Set the initial values: $\bm x^{(0)} \equiv \bm x_0$, $J^{(0)} \equiv 0$, $\bm w = \bm w_0$.
   \FOR {$i = 1$ to $I$}
  	\FOR {$i = 0$ to $N-1$}
	\STATE Compute the Euler step:
  	$$\bm x^{(k+1)} \equiv \bm x^{(k)} + \eta 
	h( \bm x^{(k)}, \tilde u(t^{(k)}), t^{(k)}).$$
	\STATE Compute the numerical integration step:
	$$J^{(k+1)} \equiv J^{(k)} + \eta F(\bm x^{(k)}, \tilde u(t^{(k)}), t^{(k)}).$$
  	\ENDFOR
	\STATE Compute the gradient of the loss function by using back propagation:
	\begin{align}\nonumber
	\bm g \equiv \nabla J(\bm w).
	\end{align}
	\STATE The weight parameter $\bm w$ is updated by using $\bm g$ (an SGD optimizer is used).
  \ENDFOR
  \STATE Output the weight vector $\bm w$.
 \end{algorithmic} 
 \end{algorithm}

In order to approximate the solution of (\ref{xevolution}), 
we have a number of numerical methods to choose from, such as Runge-Kutta methods; however, 
for simplicity, the simplest of these, the Euler method, will be used in the following argument.
Similarly, for the numerical integration of the functional (\ref{ufunctional}), 
we will exploit the simplest rectangular method.
If necessary, the accuracy of the following DU-VO method can be improved by replacing 
the ODE solver algorithm and the numerical integration method.

The core of the DU-VO method is
the Euler method for numerically solving differential equation (\ref{xevolution}),
\begin{align} \label{vo_euler}
	\bm x^{(k+1)} = \bm x^{(k)} + \eta 
	h( \bm x^{(k)}, \tilde u(t^{(k)}), t^{(k)}),\quad k =0,1,\ldots, N-1,
\end{align}
and the numerical integration step for the functional (\ref{ufunctional}),
\begin{align}\label{J_evolve}
		J^{(k+1)} = J^{(k)} + \eta F(\bm x^{(k)}, \tilde u(t^{(k)}), t^{(k)}),\quad k = 0,1, \ldots, N-1.
\end{align}

If the width of the bins is small enough, we can expect that $J^{(N)}$ 
will become an approximation of $J[\tilde u(t)]$. The concept of deep unfolding can be naturally 
applied to (\ref{vo_euler}) and (\ref{J_evolve}) to optimize the weight vector $\bm w$ 
that controls the shape of $u$.

The DU-VO method uses a loss function $J(\bm w) \equiv J^{(N)}$ 
to approximate the solution of the variational problem (\ref{vo_prob}). 
The gradient of the loss function with respect to the weight vector $\bm w$, $\nabla J(\bm w)$, 
is evaluated by back propagation. The gradient is then used to update the weight vector $\bm w$. 
This process can be optimized using common optimizers such as Stochastic Gradient Descent (SGD) or Adam. 
The entire process is summarized in Algorithm \ref{opt_alg}.

Even if the functional to be optimized contains random variables, 
the DU-VO method can minimize the expected functional ${\sf E}[J(\bm w)]$ 
by using mini-batch training. This is the one of major advantages of the DU-VO method.

\subsection{Numerical example}

To illustrate the application of the DU-VO method, 
we present a small numerical example of an optimal control problem.

The variational optimization problem is described below.
The state evolution equation is given by 
\begin{align}
	\frac{d x(t)}{dt} = u(t).
\end{align}
The functional is defined by
\begin{align}
	J[u(t)] = \int_{0}^1 (x(t)^2 + u(t)^2) dt,
\end{align}
where $u: \mathbb{R}\rightarrow \mathbb{R}$.
The boundary condition is $x(0) = 1$, and 
$x(1)$ can take any value.
The exact solution of the problem is known~\cite{keyanpour2011}: 
\begin{align} \label{exact_u}
	u^*(t) = -\frac{\sinh(1-t)}{\cosh(1)}.
\end{align}

\begin{figure}[htbp]
\begin{center}
\includegraphics[height=5cm]{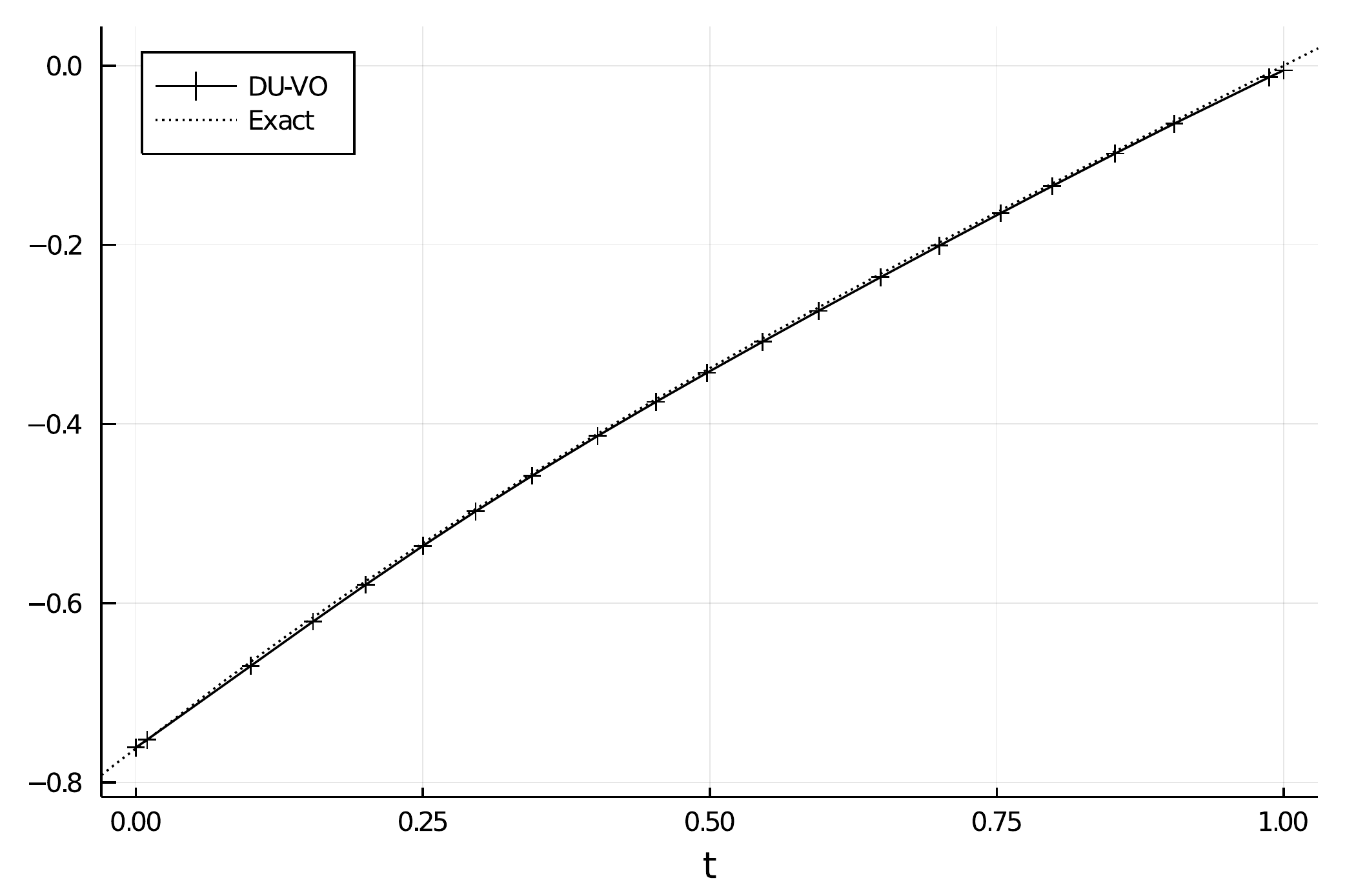}
\caption{Optimal input $u(t)$: Comparison between 
the output of the DU-VO method and the exact solution.}
\label{fig:DU-VO-example}
\end{center}
\end{figure}

Figure \ref{fig:DU-VO-example} shows the output of the DU-VO method 
and the exact solution. 
The parameter settings for the DU-VO method were as follows. The time discretization with $N = 200, T = 1$
was used. The number of optimization loops $I$ was set to 200. 
The Adam optimizer with learning rate $0.1$ was used. 
The parameter settings for the RBF approximation  
were as follows: $|c_i - c_{i+1}|=0.05, \beta = 20, S = 50$.
All the components of the weight vector $\bm w$ were initialized to 1.

It can be seen here that 
there is no obvious visible difference between the numerical results and the exact solution. 
This example thus supports the validity of the DU-VO method.

\subsection{DU-VO method for ODE-based sparse signal recovery}

In the following, we consider 
the parametric ODE (\ref{parametric_eq}) 
and the optimization problem (\ref{opt_prob}) once again.
In order to optimize the convergence behavior with respect to $\lambda(t)$,
we need a functional of $\lambda$ to be minimized that is consistent with the
optimization problem (\ref{opt_prob}).
We here introduce a functional $J$  defined by
\begin{align} \label{functional}
	J[\lambda(t)] \equiv {\sf E}\left[\int_{0}^{T^*} 
	\delta(t-T^*)\|\bm x(t) - \bm s\|^2 dt \right],
\end{align}
where $\bm s$ represents the original sparse vector and
$\bm x(t)$ in the functional is the solution of the ODE (\ref{parametric_eq}).  
The function $\delta$ represents the Dirac's delta function.
The functional (\ref{functional}) measures closeness of the solution $\bm x(T^*)$ 
to the original vector. The time $T^*$ is the predetermined target time.

The Euler recursive equation can be 
obtained by discretizing the ODE (\ref{parametric_eq}):
\begin{align} \label{du_eulier_req}
	\bm x^{(k+1)} &= \bm x^{(k)} - \eta\left(\bm A^T (\bm A \bm x^{(k)} - \bm y) + \lambda(\eta k) \tanh(\alpha \bm x^{(k)}) \right),\\
	\bm x^{(0)} &= \bm 0,
\end{align}
where $\eta = T^*/N$ is the width of the discretized bins.
The initial vector $\bm x^{(0)}$ is assumed to be the 
zero vector.
This recursive equation becomes the basis of the following DU-based optimization.
The function $\lambda: \mathbb{R} \rightarrow \mathbb{R}$
is replaced with a trainable Gaussian RBF approximation~\cite{Buhmann} defined by
\begin{align}
	\lambda(t) \equiv \sum_{i=1}^S w_i \exp( -\beta(t-\Delta i+ \theta)^2),	
\end{align}
where $\{w_i \}_{i=1}^S$ are the trainable weight parameters.
The parameters $\beta$, $\Delta$, and $\theta$ are treated as hyperparameters 
given before an optimization process. 

In an optimization process, we randomly generate 
multiple mini-batches.
A mini-batch is composed of $K$ pairs of vectors, namely 
\begin{align}
{\cal D} \equiv \{(\bm s_1, \bm y_1), \ldots, (\bm s_K, \bm y_K) \},
\end{align}
where $\bm s_i (i \in [K])$ is a sparse random vector following the Bernoulli-Gaussian distribution,
and $\bm y_i (i \in [K])$ follows our system model
$
 \bm y_i = \bm A \bm s_i + \bm n_i.
$
The noise vector $\bm n_i$ follows 
${\cal N}(\bm 0, \sigma^2 \bm I)$.

The functional $J$ can be approximated as
\begin{align}  \nonumber
	J[\lambda(t)] 
	&= {\sf E}\left[\int_{0}^{T^*} \delta(t-T^*)\|\bm x(t) - \bm x^*\|^2 dt \right] \\
	&\simeq  
	\frac 1 K \sum_{i=1}^K\|\bm x^{(N)}_i - \bm s_i\|^2,
\end{align}
where the vector 
$\bm x_i^{(N)}$ is the state vector 
corresponding to the data $(\bm s_i, \bm y_i)$. 
The state vector $\bm x_i^{(k)}$ is
evolved according to the Euler recursive equation (\ref{du_eulier_req}) 
with the initial condition $\bm x_i(0) = \bm 0$.

\begin{figure*}[htbp]
\begin{center}
\includegraphics[width=0.92\hsize]{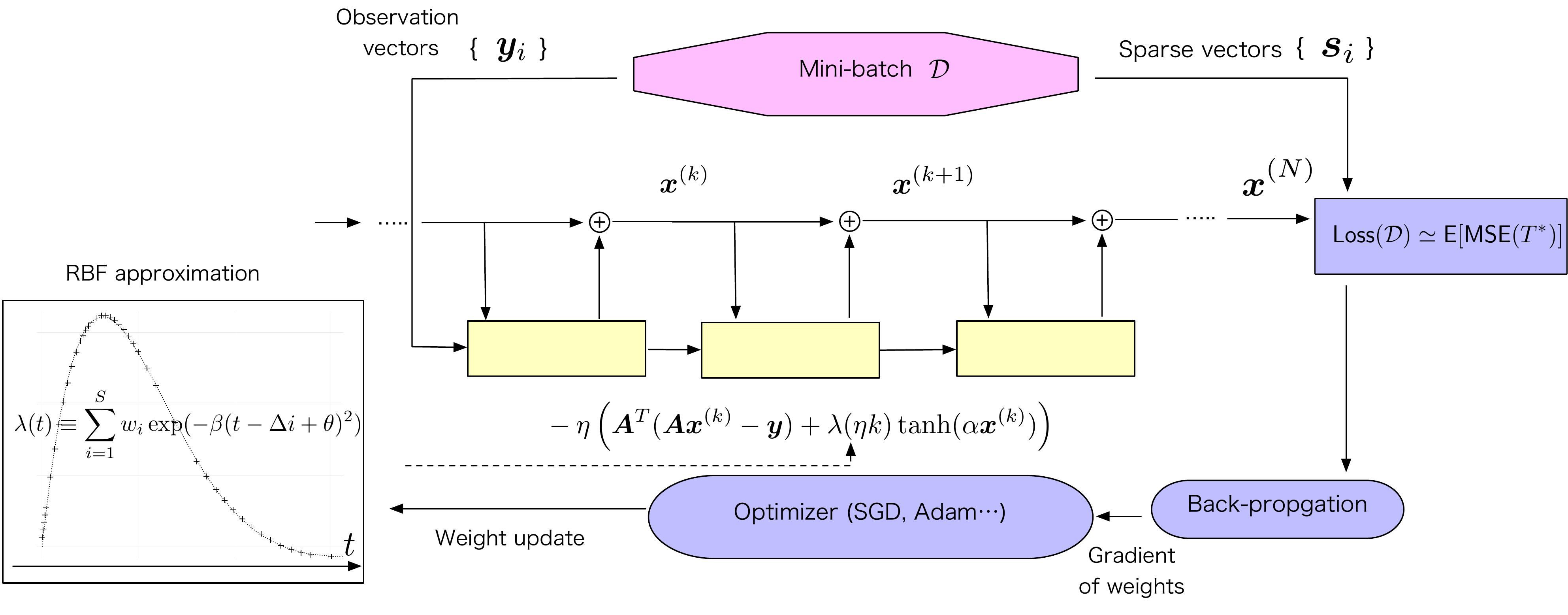}
\caption{Block diagram of Deep Unfolded-Variational Optimization (DU-VO) method 
for optimizing the parametric ODE (\ref{parametric_eq}).}
\label{DU-VO-block}
\end{center}
\end{figure*}

From this approximation above, it is reasonable 
to define the loss function: 
\begin{align}
{\sf Loss}({\cal D}) \equiv \frac 1 K \sum_{i=1}^K\|\bm x^{(N)}_i - \bm s_i\|^2,
\end{align}
which gives an approximation of MSE:
\begin{align}
{\sf Loss}({\cal D}) \simeq {\sf E}[{\sf MSE}(T^*)].
\end{align}
For each given mini-batch,  
the trainable parameters $\{w_i \}_{i=1}^S$ are updated 
based on the gradient of ${\sf Loss}({\cal D})$.
The optimization process is illustrated in 
Fig.~\ref{DU-VO-block}
 
\subsection{Numerical experiments}

\begin{figure}[htbp]
\begin{center}
\includegraphics[width=\columnwidth]{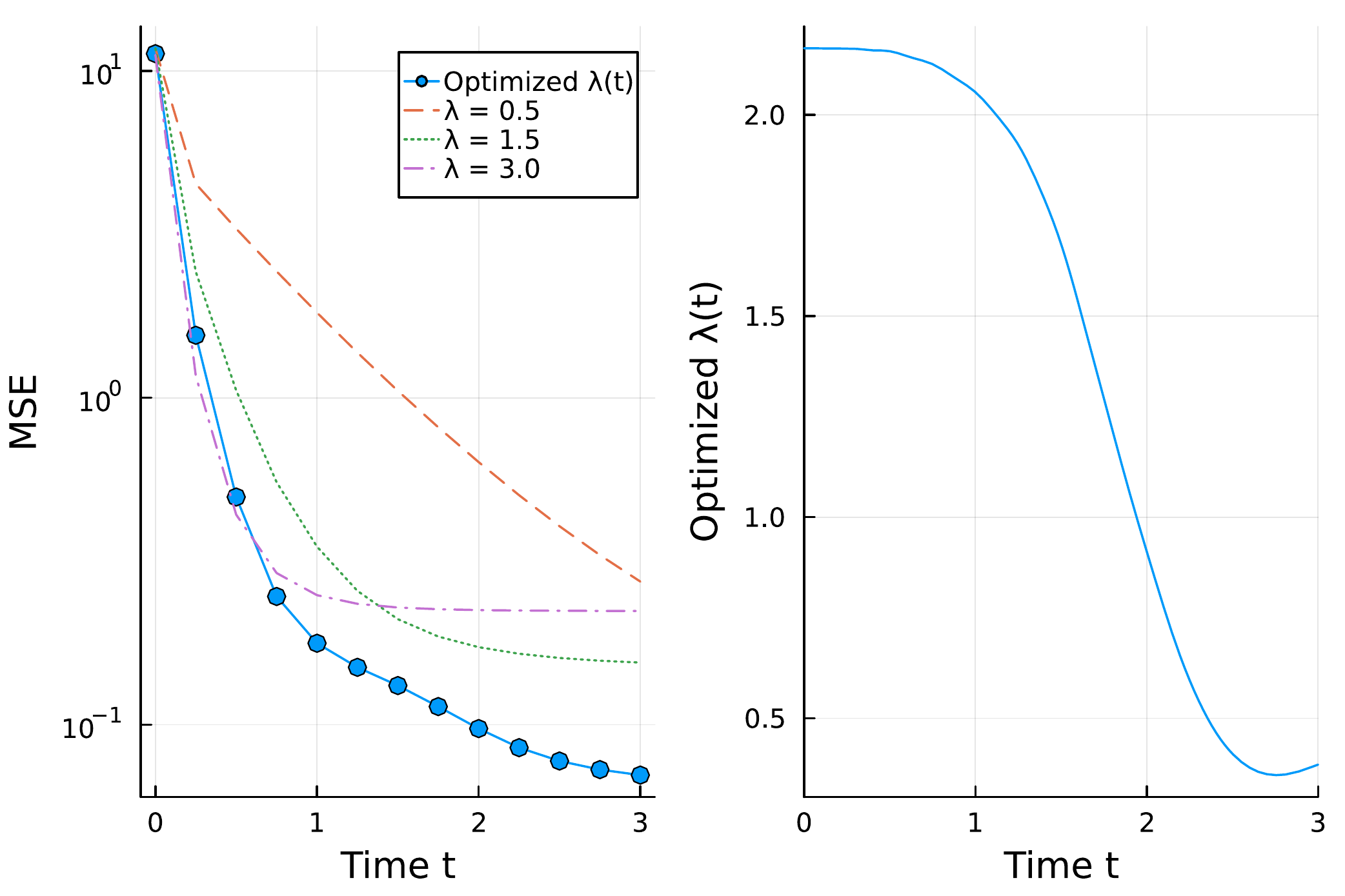}
\caption{Left: MSE as a function of $t$. Right: optimized $\lambda(t)$. Target time is $T^*=3$. 
($n=128, m = 64, p=0.1, \sigma=0.1, \alpha = 50$).}
\label{DU-VO-3}
\end{center}
\end{figure}

In the previous sections, we introduced the DU-based optimization method 
for solving variational optimization problems. 
This section presents the results of numerical experiments conducted 
to evaluate the performance of the proposed method. These experiments were performed
using the automatic differentiation mechanism provided 
by the Flux.jl library \cite{Flux} in the Julia programming language \cite{Julia}.

The problem setup for the first experiment is similar to the previous ones, 
with $n=128, m = 64, p=0.1, \sigma = 0.1$. 
The Euler method with $N = 5000$ was used for the optimization process 
and for estimating the MSE. 
The following parameters were used for the optimization process.
The mini-batch size was set to $K = 10$.
The Adam optimizer with a learning rate of $10^{-2}$ was used.
The number of mini-batches, or iterations, used for training was $100$.
The parameter settings for the RBF approximation of $\lambda(t)$ were
$\Delta = 0.25, \beta = 20, S = 20, \theta = 0.5$.

The left panel of Fig.~\ref{DU-VO-3} shows the MSE curve of the parametric ODE (\ref{parametric_eq}) with the optimized $\lambda(t)$. The MSE curves corresponding to constant regularization parameters $\lambda = 0.5, 1.5, 3.0$ are also shown. As can be seen here, the optimized $\lambda(t)$ attains the smallest MSE value among the four curves at $T^* = 3$. Additionally, the convergence rate, i.e., the slope of the optimized MSE curve, is nearly identical to that of the MSE curve with $\lambda = 0.5$ over the range $2 \le t \le 3$. The right panel of Fig.~\ref{DU-VO-3} shows the shape of the curve for the optimized $\lambda(t)$. The curve starts at a value of approximately 2.2 and rapidly decreases over the range of $1 \le t \le 2$. This result supports the argument in Subsection \ref{homotopy} and suggests that the curve of the optimized $\lambda(t)$ provides an appropriate schedule for decreasing the value of the regularization parameter.

\begin{figure}[htbp]
\begin{center}
\includegraphics[width=\columnwidth]{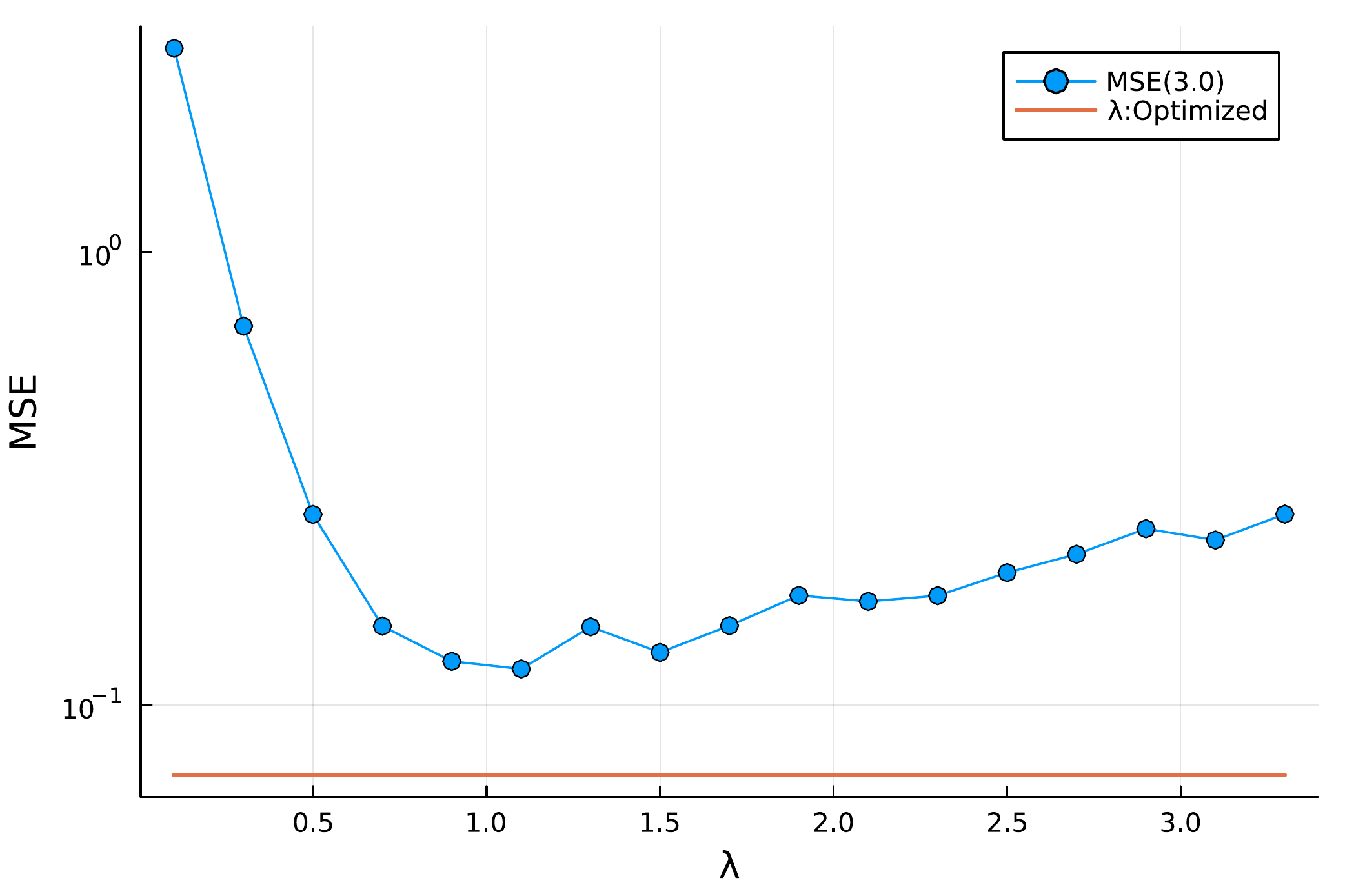}
\caption{MSE at $T^* = 3$ as a function of $\lambda$
($n=128, m = 64, p=0.1, \sigma=0.1, \alpha = 50$).}
\label{MSEat3.0}
\end{center}
\end{figure}

To confirm that the optimized $\lambda(t)$ provides the best MSE among the ODE systems with constant $\lambda$, 
the MSE at $T^* = 3$ was evaluated numerically for several values of $\lambda$. 
The results are plotted in Fig.~\ref{MSEat3.0}. As shown, the MSE values for the constant $\lambda$ 
are above $10^{-1}$, while the optimized $\lambda(t)$ achieves an MSE below $10^{-1}$. 
This result thus demonstrates the effectiveness of the proposed method in finding an optimal $\lambda(t)$ that improves the MSE.

\begin{figure}[htbp]
\begin{center}
\includegraphics[width=\columnwidth]{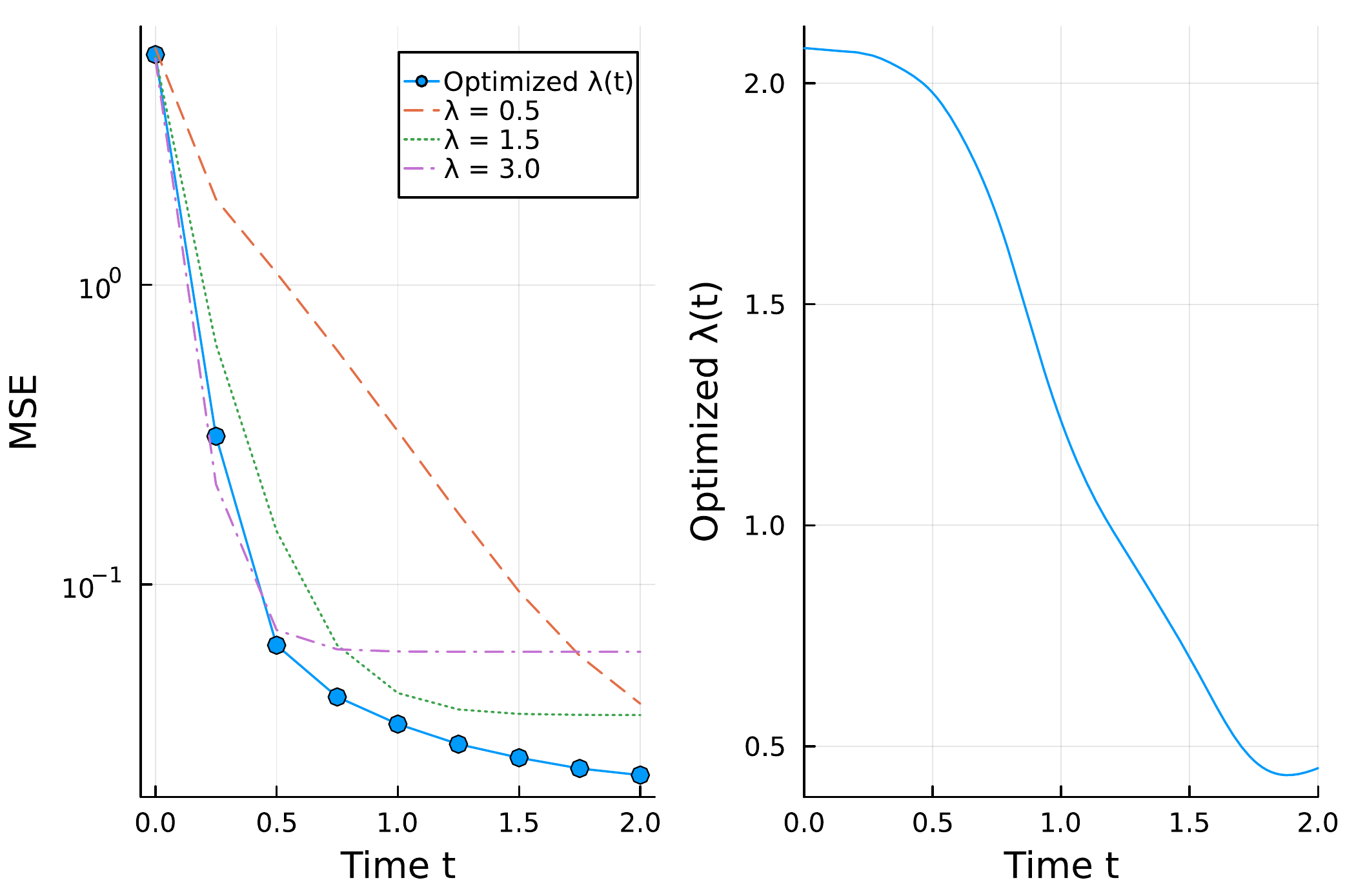}
\caption{Left: MSE as a function of $t$, Right: Optimized $\lambda(t)$.
Target time  is $T^*=2$. ($n=128, m = 64, p=0.05, \sigma=0.1, \alpha = 50$).}
\label{DU-VO-2}
\end{center}
\end{figure}

We now examine another problem setup that assumes a sparser original signal. 
Here, the probability is set to $p = 0.05$. The parameters and hyperparameters 
remain unchanged from the previous experiments, i.e., $n = 128, m = 64, \sigma = 0.1$. 
The target time is set to $T^* = 2$. The numerical results of this case are shown in Fig.~\ref{DU-VO-2}. 
It is clear that the ODE system with the optimized $\lambda(t)$ consistently gives the lowest MSE. 
It is also noteworthy that the optimized system consistently achieves close to the optimal value 
in the range of $0 \leq t \leq 2$. 
Both the convergence rate and the quality of the solution are improved by the optimization.

In summary, for both cases, the proposed DU-VO method 
successfully finds an optimized schedule for $\lambda(t)$
that improves the convergence speed and the quality 
of the solution.

\section{Conclusion}
\label{sec:conclusion}

In this study, we presented a novel approach for solving sparse signal recovery problems 
using continuous-time dynamical systems. 
The method is based on a nonlinear ODE corresponding to the gradient flow 
of the Lasso objective function. We presented the local convergence analysis of the proposed ODE 
using a linear approximation around the equilibrium point. 
In addition, we introduced a variational optimization problem to optimize
the regularization schedule and applied deep unfolding techniques \cite{LISTA, DU} 
to solve this problem. To the best of our knowledge, 
this is the first work to propose the use of deep unfolding techniques 
to solve variational optimization problems.

It should be noted that the methodology proposed in this paper can be applied 
to various forms of regularized LS problems such as binary quadratic minimization problems.
Moreover, with the future advances
in modern analog computing, especially in the optical domain, it is likely 
that ODE-based signal processing will become increasingly viable for achieving fast and energy-efficient signal processing.

\section*{Acknowledgment}
This work was supported by JSPS KAKENHI Grant Number JP22H00514.

\end{document}